%% file: ChapelierS_JCP_2017.tex
\documentclass{elsarticle}
\usepackage[T1]{fontenc}
\usepackage[latin9]{inputenc}
\usepackage{geometry}
\usepackage{color}
\makeatletter

\usepackage{lipsum}

\makeatother
\usepackage{amsmath}
\usepackage{amssymb}
\usepackage{stmaryrd}
\usepackage{graphicx}
\usepackage{graphics}
\usepackage{setspace}
\usepackage{esint}

\usepackage{multirow}
\usepackage{scalerel,stackengine}
\stackMath
\newcommand\reallywidehat[1]{%
\savestack{\tmpbox}{\stretchto{%
  \scaleto{%
    \scalerel*[\widthof{\ensuremath{#1}}]{\kern-.6pt\bigwedge\kern-.6pt}%
    {\rule[-\textheight/2]{1ex}{\textheight}}
  }{\textheight}%
}{0.5ex}}%
\stackon[1pt]{#1}{\tmpbox}%
}
\usepackage{color}
  \definecolor{darkblue}{rgb}{0.0,0.0,0.5}
  \definecolor{lightgray}{rgb}{0.95,0.95,0.95}
  \definecolor{darkred}{rgb}{0.8,0.0,0} 

\newcommand{\mr}[1]{\mathrm{#1}}

\newcommand{\f}[1]{\overline{#1}} 
\newcommand{\ff}[1]{\widetilde{#1}} 
\newcommand{\tf}[1]{\widehat{#1}} 
\newcommand{\vect}[1]{\boldsymbol{\mr{#1}}}
\newcommand{\tens}[1]{\boldsymbol{\mathsf{#1}}}

\newcommand{\pd}[2]{\frac{\partial{#1}}{\partial{#2}}} 
\newcommand{\tpd}[2]{\partial{#1}/\partial{#2}} 

\newcommand{\tred}[1]{\textcolor{black}{#1}}
\newcommand{\tblue}[1]{\textcolor{black}{#1}}

\newcommand{\methodAcronym}{CvP}
\newcommand{\methodName}{coherent-vorticity preserving~}
\newcommand{\MethodName}{Coherent-vorticity preserving~}

\graphicspath{
{./}
{./figures/filter_properties/}
{./figures/methodology/}
{./figures/numerics/}
{./figures/results/}
{./figures/results/channel_flow/}
{./figures/results/taylor_green/}
{./figures/results/helical_vortex/}
{./figures/helical_vortex/}
{./figures/visualizations/}
}

\begin{document}
{\Large{}\date{}}{\Large \par}
\begin{frontmatter}
\author[1]{J.-B. Chapelier\corref{cor1}}  
\cortext[cor1]{Corresponding author.} 
\ead{jb.chapelier@gmail.com}
\author[2]{B. Wasistho}
\address[2]{Kord Technologies, Huntsville, AL, USA}
\author[1]{and C. Scalo}
\address[1]{School of Mechanical Engineering, Purdue University, IN, USA}

\title{A \MethodName eddy-viscosity correction for Large-Eddy Simulation}


\begin{abstract}


This paper introduces a new approach to Large-Eddy Simulation (LES) where subgrid-scale (SGS) dissipation is applied proportionally to the degree of local spectral broadening, hence mitigated or deactivated in regions dominated by large-scale and/or laminar vortical motion.
The proposed \methodName (\methodAcronym) LES methodology is based on the evaluation of the ratio of the test-filtered to resolved (or grid-filtered) enstrophy, $\sigma$.
Values of $\sigma$ close to 1 indicate low sub-test-filter turbulent activity, justifying local deactivation of the SGS dissipation.
The intensity of the SGS dissipation is progressively increased for $\sigma < 1$ which corresponds to a small-scale spectral broadening.
The SGS dissipation is then fully activated in developed turbulence characterized by  $\sigma \le \sigma_{eq}$, where the value $\sigma_{eq}$ is derived assuming a Kolmogorov spectrum.
The proposed approach can be applied to any eddy-viscosity model, is algorithmically simple and computationally inexpensive.
LES of Taylor-Green vortex breakdown demonstrates that the \methodAcronym~methodology improves the performance of traditional, non-dynamic dissipative SGS models, capturing the peak of total turbulent kinetic energy dissipation during transition.
Similar accuracy is obtained by adopting Germano's dynamic procedure albeit at more than twice the computational overhead. 
A \methodAcronym-LES of a pair of unstable periodic helical vortices is shown to predict accurately the experimentally observed growth rate using coarse resolutions.
The ability of the \methodAcronym~methodology to dynamically sort the coherent, large-scale motion from the smaller, broadband scales during transition is demonstrated via flow visualizations.
Finally, the CvP methodology is also shown to improve the accuracy of subrid models in LES of compressible turbulent channel flow. 
\end{abstract}


\end{frontmatter}

\input{introduction.tex}
\input{numerics.tex}
\input{methodology.tex}
\input{results.tex}


\section{Conclusion}

In this paper, a novel SGS modeling concept based on a turbulence sensor has been developed.
The turbulence sensor is build from the ratio of test and grid filtered esntrophy in order to determine the presence of transitional or broadband turbulence locally in the flow.
A damping function depending on the sensor is defined and allows for a reduction of the modeled subgrid dissipation in laminar or transitional regions.

The present methodology has first been assessed from LES computations of the Taylor-Green vortex flow at $Re=5000$.
It is found that the damping function successfully reduces the SGS dissipation at the early, transitional stages of the flow and progressively leads to a full activation of the subgrid model when turbulence is building up in the flow.
The corresponding evolution of kinetic energy and dissipation is very well predicted and matches closely filtered DNS results, especially considering the coarse discretization of the LES study. 
The damping function has been coupled to several dissipative LES models and is found to improve all of them, underlying the flexibility of the method which can be used in conjunction with any existing SGS model.
Different test filters for the computation of the sensor function have been assessed and a mild variation of the results is observed.
This is a very encouraging results for the extension of the present approach to other type of numerical methods, for which the freedom of choice for the test filter can be restricted due to implementation difficulties or stencil limitations.
The method is also robust regarding the discretization employed as good results are reported for different grid sizes.
It must be emphasized that this approach is relatively simple, inexpensive and only the implementation of a test filter is required.

The CvP approach has been assessed as well for a newly developed numerical test case allowing for the fundamental study of rotor wake vortices.
\tblue{LES computations} of this configuration with a pitch to vortex radius ratio $h/R=1.1$, which is unstable due to the mutual inductance phenomenon, \tblue{are} found to accurately predict the growth rate observed in the experiments.
Visualizations of iso-surfaces of vorticity colored by the sensor function values have shown that the present turbulence sensor is able to sort the coherent motion from the small-scale turbulence.

Finally, the CvP approach has been evaluated in the context of wall-bounded turbulence. 
\tblue{The near-wall values of the CvP sensor have been studied a priori from DNS data, and it is found that the sensor provides the correct near-wall scaling when the enstrophy is not test-filtered in the spanwise direction. This study paves the way towards the development of modified versions of CvP that account for flow anisotropy.}
\tblue{LES of the channel flow have been conducted considering the Vreman model and its CvP variant, and the CvP-Vreman computation yields a good agreement with DNS and is found to improve the mean and fluctuating velocity profiles compared to the baseline model.}

Future work will aim at using the presently developed methodology to explore the physics of various configurations of helical vortices at high Reynolds number for which DNS resolutions are unattainable.
The CvP methodology will also be used to understand the dynamics of vortex entanglement via simulations of knotted vortex configurations~\cite{kleckner2013creation}.

\tblue{The CvP approach could also be considered in the context of non-eddy viscosity subgrid models, such as scale-similarity operators that provide an explicit evaluation of the subgrid scale field.}

\section*{Acknowledgments}

The authors acknowledge financial support of the subcontract KSC-17-001 between Purdue University and Kord Technologies, Inc (Huntsville), conducting research under the US Navy Contract N68335-17-C-0159 STTR-Phase II, Purdue Proposal No. 00065007, Topic N15A-T002 entitled \emph{Vortex Preserving and Consistent Large Eddy Simulations for Naval Applications}. Computational resources where provided by the Rosen Center for Advanced Computing (RCAC) at Purdue University and Information Technology at Purdue (ITaP).

\section*{References}

\bibliographystyle{aiaa}
\bibliography{references}

\end{document}

%% file: introduction.tex

\section{Introduction}

\tred{The Large-Eddy Simulation (LES) technique is a tool of prime importance as it enables an accurate numerical prediction of high-Reynolds-number flows of practical relevance at accessible computational costs. In particular, the explicit representation of the spatial and temporal dynamics of vortices makes LES more accurate and versatile than Reynolds-Averaged Navier-Stokes (RANS) approaches, while its reduced requirements in terms of degrees of freedom compared to Direct Numerical Simulations (DNS) allows for tackling high-Reynolds-number flows.} The development of accurate and robust sub-grid scale (SGS) models extending the envelope of attainable Reynolds numbers, while containing the computational cost, is therefore still a warranted effort.

One shortcoming of traditional LES modeling approaches is their tendency to introduce excessive SGS dissipation in transitional regions, impacting the evolution of the large coherent structures which may be on the verge of break-up.
The Smagorinsky model~\cite{smagorinsky1963general}, for example, attenuates velocity gradients at all scales of the flow, resulting in the undesired damping of coherent laminar vortices or transitional regions.
Any accurate SGS model should therefore account for the energy transfers towards unresolved scales that are concentrated in the spectral neighborhood of the cutoff wavelength \cite{kraichnan1976eddy,sagautLESbook}

A number of approaches aim at correcting the overly dissipative nature of SGS models, and improving their spectral properties.
The so-called Dynamic procedure adjusts automatically the subgrid model parameter, modulating the intensity of the subgrid dissipation for transitional or inhomogeneous flows~\cite{germano1991dynamic,ghosal1995dynamic,meneveau1996lagrangian,salvetti1995priori,lamballais1998spectral,bou2005scale,park2006dynamic}.
Alternate multiscale approaches aim at applying a high-pass filter on the subgrid tensor to reduce the influence of the subgrid dissipation on the energy-containing large scales.
\tred{The Variational Multiscale (VMS) approach introduced by Hughes et al.~\cite{hughesLESVMS} and further developed in other works~\cite{hugheschannelflow,koobus2004variational,farhat2006dynamic,bricteux2009multiscale,wasberg2009variational,ouvrard2010classical,meyers2007evaluation,chapelier2016development}, aims at building a small-scale dissipative operator by performing an explicit scale-separation of the strain rate tensor. Among related approaches and variants, Stolz et al.~\cite{stolz2005high} proposed to apply a high-pass filtered on the eddy viscosity, Vreman~\cite{vreman2003filtering} applied a high-pass filter on the whole subgrid dissipative operator while Jeanmart and Winckelmans~\cite{jeanmart2007investigation} applied the high-pass filter to the strain rate tensor. All the aforementioned approaches lead to significant improvement of the subgrid models accuracy.}
Other authors have proposed the use of turbulence sensors attempting to discriminate between laminar and turbulent regions~\cite{david1993modelisation,ackermann2001modified,chapelier2016spectral}.
Another approach termed Coherent Vortex Simulation (CVS) uses a wavelet-based decomposition to sort the coherent motion from the Gaussian component of the solution identified as small-scale noise~\cite{farge1999non,farge2001coherent,schneider2005coherent,farge2001coherent2}, adopting a signal-processing and statistical approach. \tblue{Alternative approaches aim at removing the subfilter energy by devising minimum-dissipation eddy-viscosity-based SGS models~\cite{rozema2015minimum}.}

Purely numerical approaches have been proposed and rely on the use of regularization procedures such as explicit filtering or addition of artificial viscosity to counter the high-wave-number-energy accumulation occurring in low-dissipation numerical schemes~\cite{karamanos2000spectral,pasquetti2006spectral,bogey2006large,bogey2006computation,dairay2017numerical}.

Other approaches, termed Implicit LES (ILES), aim at tailoring the numerical dissipation naturally present in the adopted numerical discretization, to mimic the sub-grid dissipation deriving from physical SGS models~\cite{grinstein2007implicit,thornber2007implicit,hickel2006adaptive,hickel2007implicit}.
Although resulting in almost no computational overhead, as no explicit execution of subgrid models is needed, such techniques may still introduce excessive dissipation~\cite{garnier1999use,mittal1997suitability}, with less flexibility than inherently non-dissipative numerical schemes equipped with subgrid models, which can be deactivated in dynamically selected regions of the flow.
The latter will be the approach followed in this study.

In fact, accurate results have been obtained in the past by coupling high-order finite difference schemes and Dynamic models~\cite{nagarajan2003robust,kravchenko1997effect,morinishi1998fully}.
Classic Dynamic modeling approaches, however, increase cost, memory requirements and complexity of the implementation, associated with the test-filtering of tensors and averaging of the dynamic parameter along directions of statistical homogeneity \cite{germano1991dynamic} or flow path trajectories \cite{meneveau1996lagrangian} to obtain stable computations.

In this paper, we present a new strategy for quantifying the local degree of spectral broadening in the flow with a simple and computationally inexpensive turbulence sensor identifying regions of developed, locally high Reynolds number turbulence requiring SGS dissipation. 
This approach can be seen as a new dynamic approach, blending physics-based SGS modeling with a new scale-selective sensor based on the evaluation of the sub-test-filter enstrophy content.
By test-filtering vorticity rather than velocity, a greater sensitivity to the emergence of small-scales in the flow is achieved.
\tblue{Indeed, while the bulk of the energy in the flow is carried by the large-scale motion, the small scales are characterized by high levels of enstrophy.
Hence, defining a scale separation of enstrophy is a natural choice for the development of a sensor detecting small-scale dynamics.}
By evaluating the ratio of grid- and test-filtered enstrophy, it is possible to quantify the relative small-scale energy content of the flow to then mitigate SGS dissipation in non-turbulent, large-scale narrowband vorticity dominated regions where coherent or large-scale vortices are likely to be found.
The sensor is hence able to discriminate between gradients due to small scale vorticity (most likely broadband turbulence) and large scale structures, which are not governed by inertial subrange transfer energy dynamics assumed by most SGS models. \tblue{The subgrid dissipation will therefore be activated only in regions where under-resolved, small-scale turbulence is prominent.} Due to this property we refer to the proposed method as~\methodName.

The present approach has a number of advantages: (1) the sensor improves the accuracy of any existing SGS model in transitional flows; (2) the sensor is based on local and instantaneous flow values, allowing for a dynamic adjustment of the SGS dissipation; (3) the \methodAcronym~technique is computationally inexpensive, only requiring one addition test-filtering operation on the enstrophy field; (4) algorithmically simple, not requiring any spectral decomposition of the flow and easily extendable to unstructured meshes.

\tred{As advanced subgrid modeling approaches such as the Dynamic model or the VMS approaches provide as well a good accuracy for transitional and complex flows, the CvP methodology is expected to be computationally less intensive. Indeed, the VMS requires the scale separation of all velocity gradients in order to build the small-scale strain-rate tensor and the Dynamic approach requires the test filtering of numerous quantities for computing the dynamic parameter. The CvP-LES approach requires the filtering of only one quantity, the enstrophy, yielding a minimal computational overhead.}

The outline of the paper is as follows.
First, the LES formalism and numerical methods are detailed in section \ref{sec:numerics}.
The \methodAcronym~technique is derived in section \ref{sec:cvp_method}.
The various subgrid-scale models used in the study are reported in section~\ref{sec:numerics:sgs_models}.
Section \ref{sec:results_TGV} features a sensitivity study of the~\methodAcronym~technique  to the grid resolution and test filter width in LES of transitional Taylor-Green vortex.
The sensor is coupled to various traditional dissipative SGS models, and compared to the classical Dynamic model. \tblue{It is found that the CvP-LES approach improves significantly the accuracy of the baseline SGS models and competes favourably against the Dynamic model in terms of accuracy, for a reduced computational cost. The CvP method is also found to be robust to variations in the type of test filter chosen.}
The approach is then applied to a periodic double helical vortex configuration (section \ref{sec:results:helical_vortex}), which is a model problem for rotor blade vortices.
The ability of the sensor function to sort the coherent structures from the broadband, small-scale turbulence is also assessed with instantaneous visualizations. \tblue{The method is also found to yield an accurate description of the vortex core deviation rates as shown from a comparison with experiments.}
\tblue{Section~\ref{sec:results:channel} features the development and evaluation of the CvP approach for wall-bounded turbulence. An a priori study of the CvP sensor function is conducted from a DNS of the channel flow at bulk Reynolds number 3000 and Mach number 1.5, and a CvP-LES computation of this flow is conducted, yielding a good agreement with DNS.} 

\tblue{In the last section, the main findings of the paper are summarized and future directions are discussed.}

%% file: numerics.tex

\section{Numerics} \label{sec:numerics}

\subsection{Filtered Navier-Stokes equations} \label{sec:numerics:ns}

In this work, the compressible fluid motion is simulated by discretizing the Navier-Stokes operator $\mathcal{NS}\left(\vect{w}\right)$, which can be cast in the form:
\begin{equation}
\mathcal{NS}(\vect{w})=\pd{\vect{w}}{t}+\vect{\nabla}\cdot\left[\tens{F}_{\mr{c}}(\vect{w})-\tens{F}_{\mr{v}}(\vect{w},\nabla\vect{w})\right]=\vect{0},
\end{equation}
where $\vect{w}=\left(\rho,\rho\vect{U},\rho E\right)^{\mr{T}}$ is the vector of conserved variables $\rho$, $\vect{U}$ and $E$, density, velocity and total energy respectively, and $(\nabla\vect{w})_{ij} = \tpd{w_i}{x_j}$ its gradient.
The viscous and convective flux tensors $\tens{F}_{\mr{c}},\tens{F}_{\mr{v}}\in\mathbb{R}^{5\times3}$ read 
\begin{equation}
\tens{F}_{\mr{c}} =
\begin{pmatrix}
\rho\vect{U}^{\mr{T}}\\
\rho\vect{U}\otimes\vect{U}+ p\tens{I}\\
(\rho E+p )\vect{U}^{\mr{T}}
\end{pmatrix},\quad\text{and}\quad 
\tens{F}_{\mr{v}} = 
\begin{pmatrix}
\vect{0}\\
\tens{\tau}\\
\tens{\tau}\cdot\vect{U}-\lambda\vect{\nabla} T^{\mr{T}}
\end{pmatrix},
\end{equation}
where $T$ is the temperature, $p$ is the pressure, $\lambda$ is the thermal conductivity of the fluid and $\tens{I} \in \mathbb{R}^{3\times3}$ is the identity matrix.
For a Newtonian fluid, we have 
\begin{equation}
\tens{\tau}=2\mu\tens{S},
\end{equation}
where $\mu$ is the dynamic viscosity and 
\begin{equation}
\tens{S}=\frac{1}{2}\left[\nabla\vect{U}+\nabla\vect{U}^{\mr{T}}-\frac{2}{3}\left(\vect{\nabla}\cdot\vect{U}\right)\tens{I}\right]
\label{eq:shear:stress}
\end{equation}
is the strain rate tensor.
The ideal gas law is considered for the closure of the system of equations, namely,
\begin{equation}
p=(\gamma-1)\left(\rho E-\frac{1}{2}\rho\vect{U}\cdot\vect{U}\right),
\end{equation}
where $\gamma$ is the heat capacity ratio.

The LES equations are obtained by applying a low-pass filter to the Navier-Stokes equations~\cite{leonard1974energy}.
The spatial filtering operator applied to a generic quantity $\phi$ reads 
\begin{equation}
\f{\phi}(\vect{x},t)=g (\vect{x})\star\phi (\vect{x}),
\label{eq:filt}
\end{equation}
where $\star$ is the convolution product and $g\left(\vect{x}\right)$ is a filter kernel related to a cutoff length scale $\overline{\Delta}$ in physical space~\cite{sagautLESbook}. The compressible case requires density-weighted filtering approaches. The density-weighted or Favre filtering operator is defined as
\begin{equation}
\ff{\phi}=\frac{\f{\rho\phi}}{\f{\rho}}.
\end{equation}

In the present study, the compressible LES formalism introduced by Lesieur et al.~\cite{lesieur2001favre,lesieur2005large,lesieur1996new} is adopted yielding the following set of filtered compressible Navier-Stokes equations:
\begin{equation}
\mathcal{NS}(\f{\vect{w}}) = \vect{\nabla} \cdot \tens{F}_{\mr{SGS}}(\f{\vect{w}},\nabla\f{\vect{w}}),
\label{eq:filtns}
\end{equation}
where $\f{\vect{w}}=\left(\f{\rho},\f{\rho}\ff{\vect{U}},\f{\rho}\ff{E}\right)^{\mr{T}}$ is the vector of filtered conservative variables.

The SGS tensor $\tens{F}_{\mr{SGS}}$ is the result of the filtering operation and it encapsulates the dynamics of the unresolved sub-grid scales, and is modeled here using the eddy-viscosity assumption: 
\begin{equation}
\tens{F}_{\mr{SGS}}(\f{\vect{w}},\nabla\f{\vect{w}}) = 
\begin{pmatrix}
\vect{0}\\
2\mu_{t}\f{\tens{S}}\\
-\frac{\mu_{t}C_{p}}{Pr_{t}}\vect{\nabla}\ff{T}^{\mr{T}}
\end{pmatrix},
\end{equation}
where $\f{\tens{S}}$ is the shear stress tensor computed from equation~\eqref{eq:shear:stress} based on the Favre-filtered velocity $\ff{\vect{U}}$, $Pr_{t}$ is the turbulent Prandtl number, which is set to 0.5~\cite{erlebacher1992toward}, $C_{p}$ is the heat capacity at constant pressure of the fluid and $\mu_{t}$ is the eddy-viscosity which depends on the chosen sub-grid model.

The only assumption required to derive~\eqref{eq:filtns} is that the filtering operation~\eqref{eq:filt} commutes with spatial derivatives.
The various filtering and discretization strategies adopted in the current work are outlined in the following sections.

\subsection{Numerical Discretization} \label{sec:numerics:numerics}

The compressible, Favre-filtered Navier-Stokes equations are solved using a sixth order compact finite difference scheme solver originally written by Nagarajan \emph{et al.}~\cite{nagarajan2003robust}, currently under development at Purdue University.
The solver is based on the staggered grid arrangement illustrated in Figure 1, providing superior accuracy compared to a fully collocated approach~\cite{lele1992compact}.

\begin{figure}[!h]
\centering
\includegraphics[width=0.95\linewidth]{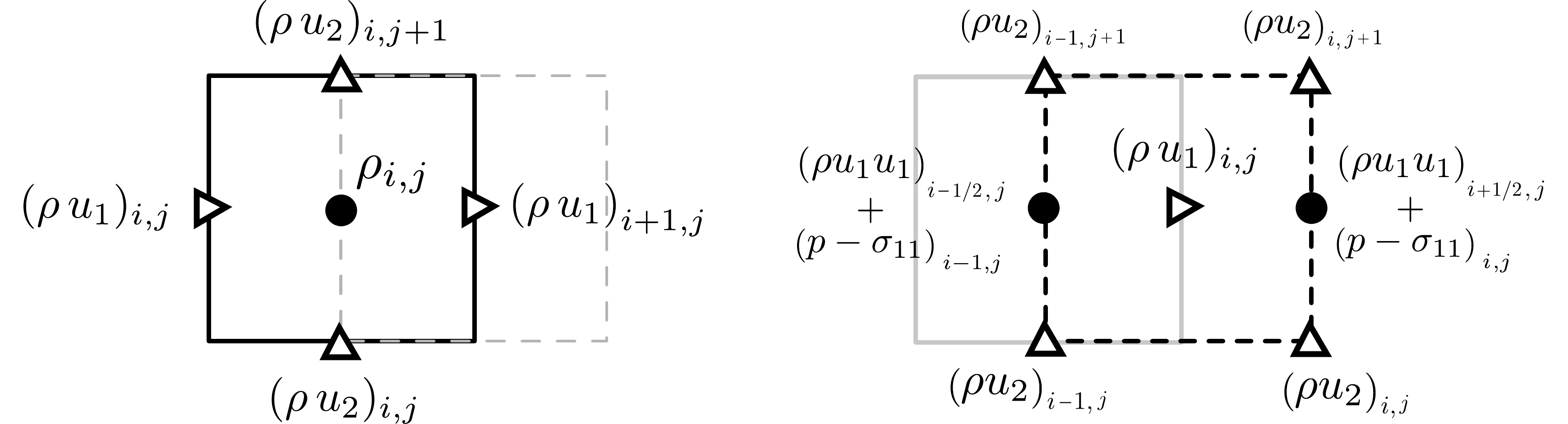}
\caption{Variables location on a staggered grid for the continuity (left) and one component of the momentum (right) equation.}
\label{staggered}
\end{figure}
The time integration is performed using a third order Runge-Kutta scheme.

\subsection{Filtering Strategy}  \label{sec:numerics:filters} \label{sec:testfilters}

In the present study, three different test filters with different filter lengths are considered for the purpose of assessing the influence of the test filter on the new model's performance.
The generic formulation for the computation of the test-filtered quantities reads:
\begin{equation}
\alpha \widehat{\overline{f}}_{i-1}+\widehat{\overline{f}}_i+\alpha \widehat{\overline{f}}_{i+1}=a\overline{f}_i+\frac{b}{2}(\overline{f}_{i+1}+\overline{f}_{i-1})+\frac{c}{2}(\overline{f}_{i+2}+\overline{f}_{i-2})+\frac{d}{2}(\overline{f}_{i+3}+\overline{f}_{i-3})+\frac{e}{2}(\overline{f}_{i+3}+\overline{f}_{i-3}),
\end{equation}

\begin{figure}[!h]
\centering
\includegraphics[width=0.85\linewidth]{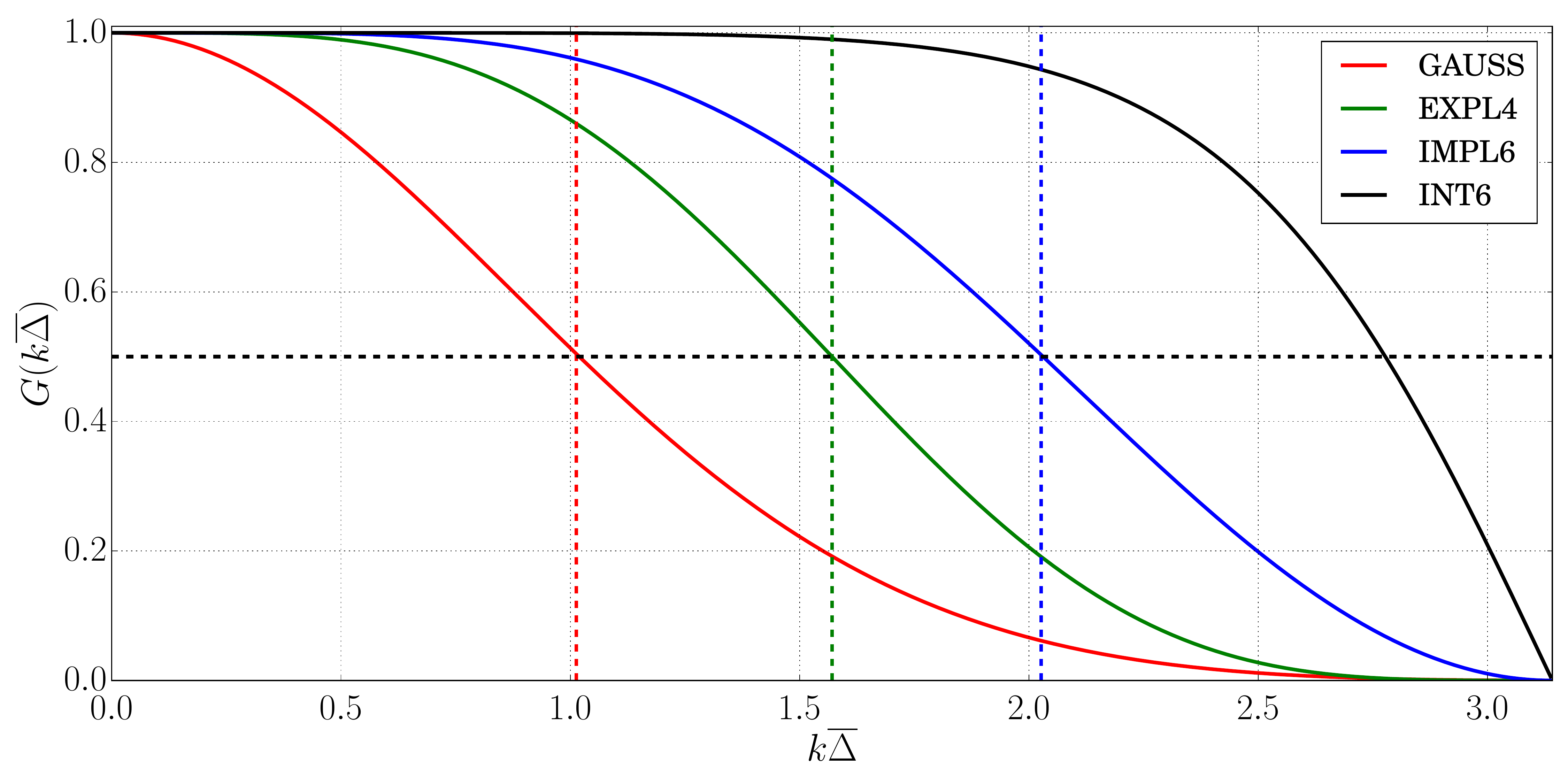}
\caption{Transfer functions of the various test filters in Table 1 and interpolation operators required by the staggered grid approach.}
\label{test_filters_tf}
\end{figure}

The first test filter considered is a sixth-order, spatially implicit compact filter introduced by Lele~\cite{lele1992compact}.
This filter is characterized by non-zero values of the parameter $\alpha\in]-0.5,0.5[$, which controls the strength of the test filter.
Weaker filters are obtained using high values of $\alpha$ and vice-versa.
This filter is named IMPL6 for the rest of the paper.
Two additional explicit filters with larger filter widths are considered:
a fourth-order explicit test filter (EXPL4) and the Gaussian test filter (GAUSS) proposed by Cook and Cabot~\cite{cook2005hyperviscosity}). 
The finite difference coefficients corresponding to each test filter are detailed in Table~\ref{fdcoeff_tf}.
The transfer functions $G(k\overline{\Delta})$ of the test filters considered here are plotted in Figure~\ref{test_filters_tf}.
GAUSS is the strongest filter, while EXPL4 is intermediate and the IMPL6 with $\alpha$ set to -0.4 the weakest.
Their filter length are respectively $\widehat{\overline{\Delta}}_{\mathrm{GAUSS}}=3\overline{\Delta}$, $\widehat{\overline{\Delta}}_{\mathrm{EXPL4}}=2\overline{\Delta}$ and $\widehat{\overline{\Delta}}_{\mathrm{IMPL6}}=1.5\overline{\Delta}$, as defined by the cutoff wavenumber verifying $G(\widehat{\bar{k}} \overline{\Delta})=0.5$, where $\widehat{\overline{k}}=\pi/\ \widehat{\overline{\Delta}}$.

{\def\arraystretch{2}\tabcolsep=5pt
\begin{table}[t]
\centering
\begin{tabular}{c|ccccccc}
Filter & $\widehat{\overline{\Delta}}$ & $\alpha$ & a & b & c & d & e  \tabularnewline
\hline 
IMPL6 & $1.5\overline{\Delta}$ & $\alpha$ & $\frac{1}{16}(11+10\alpha)$    & $\frac{1}{32}(15+34\alpha)$ & $\frac{1}{16}(-3+6\alpha)$ &  $\frac{1}{32}(1-2\alpha)$ & $0$ \tabularnewline
EXPL4 & $2.0\overline{\Delta}$ & $0$      & $\frac{1}{2}$ & $\frac{9}{16}$ & $0$ & $-\frac{1}{16}$ & $0$ \tabularnewline
GAUSS & $3.0\overline{\Delta}$ & $0$      & $\frac{3565}{10368}$ & $\frac{3091}{12960}$ & $\frac{1997}{25920}$ & $\frac{149}{12960}$ & $\frac{107}{103680}$ \tabularnewline
\end{tabular}
\caption{Finite difference coefficients for the LES test filters.}
\label{fdcoeff_tf}
\end{table}
}
The computation of the turbulence sensor requires the interpolation of the velocity gradients from their natural locations to the $\rho$ locations, which is performed using a sixth order finite difference interpolation (INT6).
The transfer function of the interpolant is accounted for in the development of the CvP-LES methodology.

%% file: methodology.tex

\section{\MethodName (\methodAcronym) Eddy-Viscosity Correction} \label{sec:cvp_method}
\subsection{Construction of the CvP sensor}
In this section, the details of the CvP-LES methodology are presented.
The CvP sensor, which is able to discriminate between coherent and broadband turbulence, is based on the detection of vorticity in the range of scales located between the primary and test filter cutoff lengths.
The primary filtering implicitly performed by the computational grid and its cutoff length is $\overline{\Delta}=\Delta$, which is the cell size.
The test-filter is designed to isolate large scales and its cutoff length $\widehat{\overline{\Delta}}$ is therefore greater than $\overline{\Delta}$.

\begin{figure}[!h]
\centering
\includegraphics[width=0.98\linewidth]{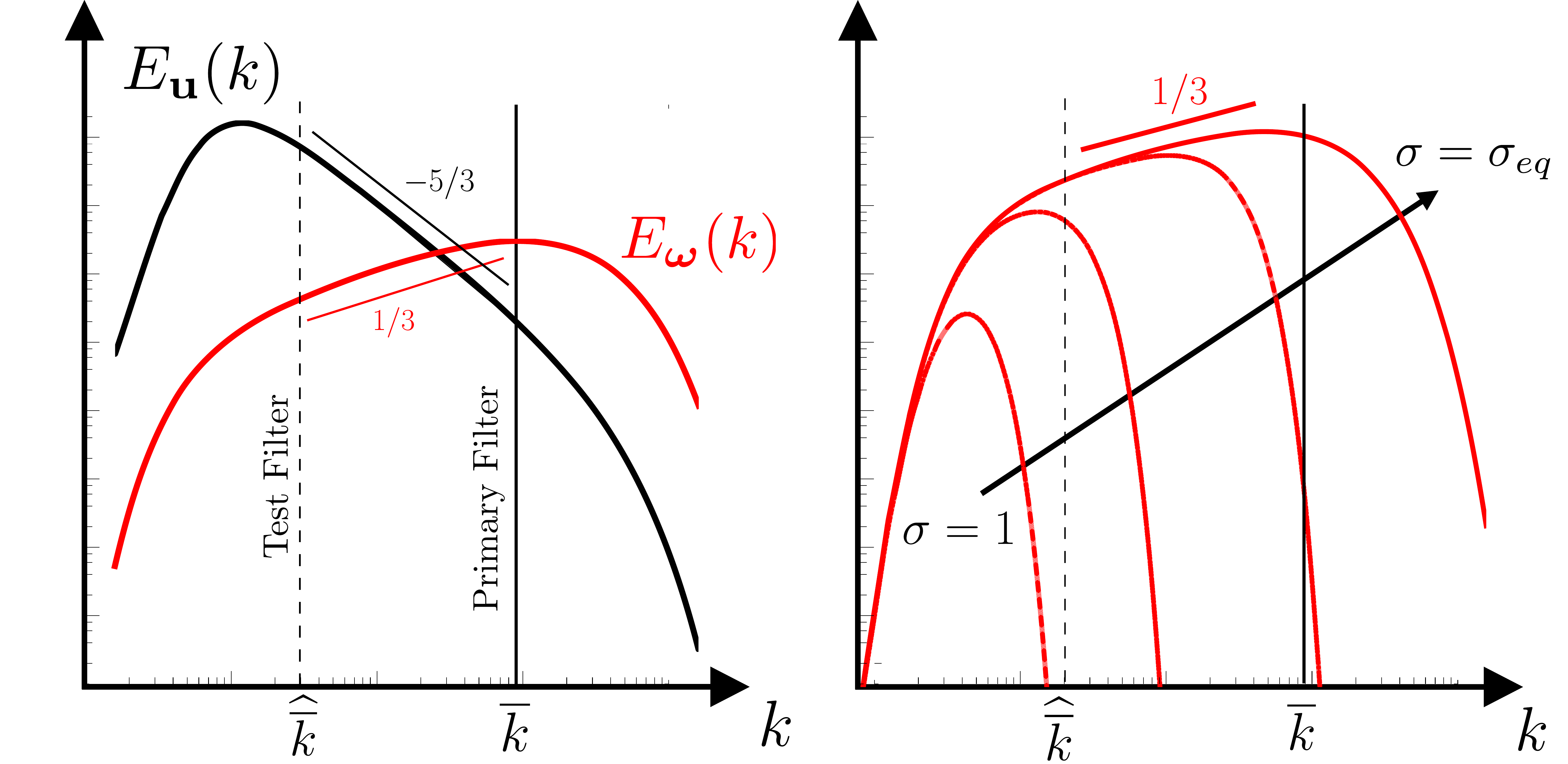}
\caption{Left Figure: Qualitative depiction of energy (black) and enstrophy (red) spectra for high-Reynolds number isotropic turbulence.
Right Figure: Qualitative evolution of enstrophy spectra in transitional flows from $\sigma=1$ (initial condition) to $\sigma=\sigma_{eq}$ (developed turbulence).}
\label{model_spectra}
\end{figure}

In idealized incompressible, equilibrium isotropic turbulence, the transfer of energy, $E=\boldsymbol{\mathrm{u}}\cdot\boldsymbol{\mathrm{u}}$/2, from the large scales towards the small scales of the flow, yields a broadband energy spectrum scaling as $k^{-5/3}$ in the inertial range.
This particular distribution of the energy in the spectral space implies that most of the energetic content of the flow is carried by the large scales.
On the contrary, the spectral content of enstrophy $\xi=\boldsymbol{\omega}\cdot\boldsymbol{\omega}/2$, where $\boldsymbol{\omega}$ is the vorticity vector,  grows as $k^{1/3}$ in the inertial range, peaking at the small scales. 
In developed turbulent flows, the enstrophy levels are therefore mostly governed by the small-scale activity.
These observations suggest that a small-scale activity sensor can be built by comparing the test-filtered and resolved enstrophy via the ratio:
\begin{equation}
\sigma=\frac{\tf{\f{\xi}}}{\: \overline{\xi} \:},
\end{equation}
where $\tf{\f{\xi}}=\frac{1}{2}\tf{\f{\boldsymbol{\omega}}\cdot\f{\boldsymbol{\omega}}}$ is the test-filtered enstrophy and $\f{\xi}=\frac{1}{2}\f{\boldsymbol{\omega}}\cdot\f{\boldsymbol{\omega}}$ the resolved enstrophy.

The sensor is built based on the following simple observations:
\begin{itemize}
\item If $\hat{\bar{\xi}}\approx\bar{\xi}$, no significant vorticity is detected at the small scales and subgrid dissipation should be deactivated.
This corresponds to values of $\sigma$ close to 1 indicating low sub-test-filter turbulent activity.
\item If  $\hat{\bar{\xi}}\ll\bar{\xi}$, a sub-test-filter vortical activity is detected and subgrid dissipation should be applied proportionally to the degree of spectral broadening.
This corresponds to values of $\sigma < 1$, which span conditions ranging from incipient spectral broadening to equilibrium turbulence. 
\end{itemize}
The next step consists in creating a function $f(\sigma)$, admitting variations between 0 and 1, which will act as a turbulence sensor.
An adaptation of the subgrid model intensity is achieved by multiplying the eddy viscosity by the sensor function:
\begin{equation}
\mu_t^{\mathrm{CvP}}=f(\sigma)\mu_t
\label{eq:cvp_mut}
\end{equation}
This application of the sensor via equation~(\ref{eq:cvp_mut}) leads to a decrease of the eddy viscosity in transitional and smooth regions of the flow and a progressive increase depending on the extent of the local small-scale enstrophy content.

\begin{figure}[!h]
\centering
\includegraphics[width=0.49\linewidth]{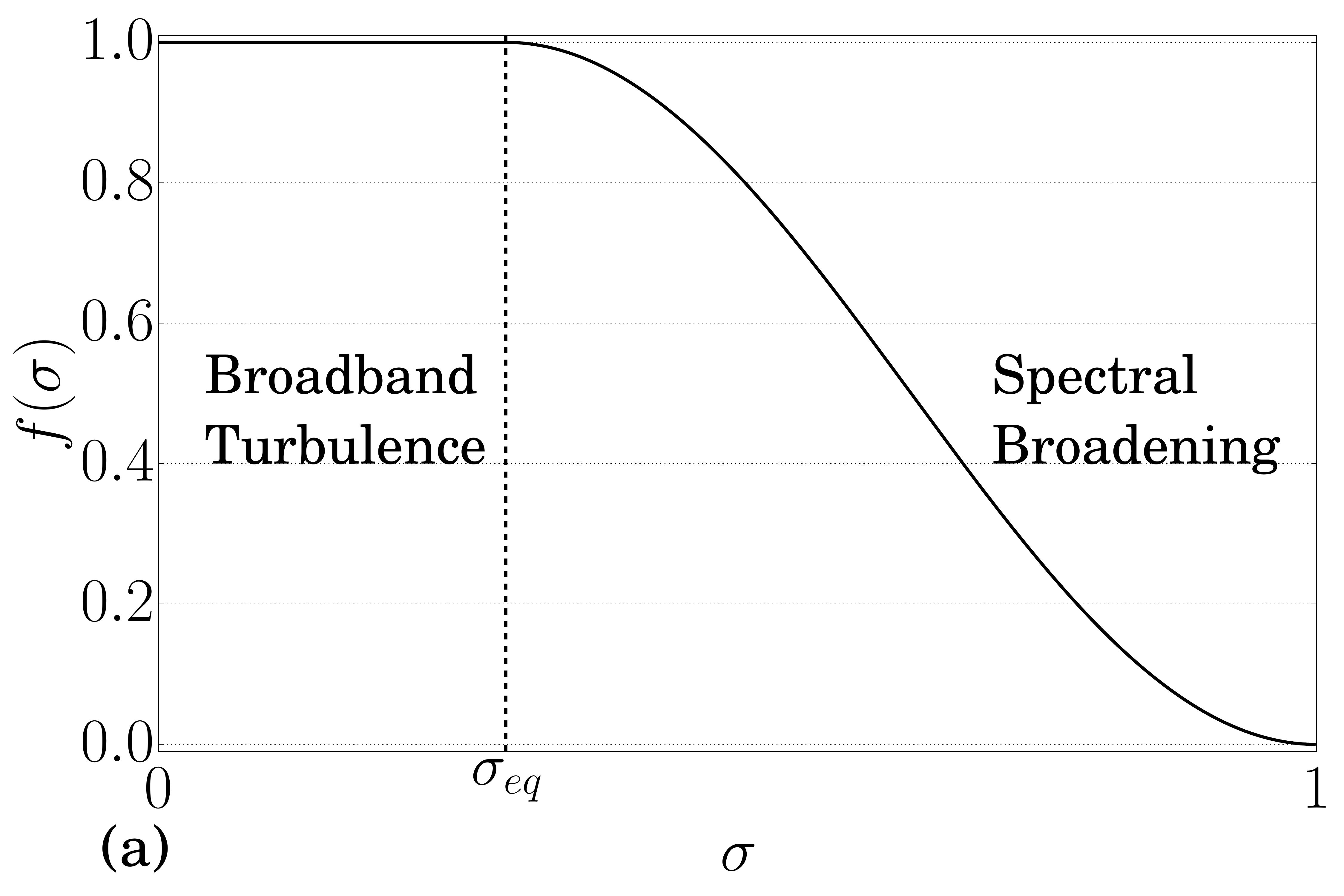}
\includegraphics[width=0.49\linewidth]{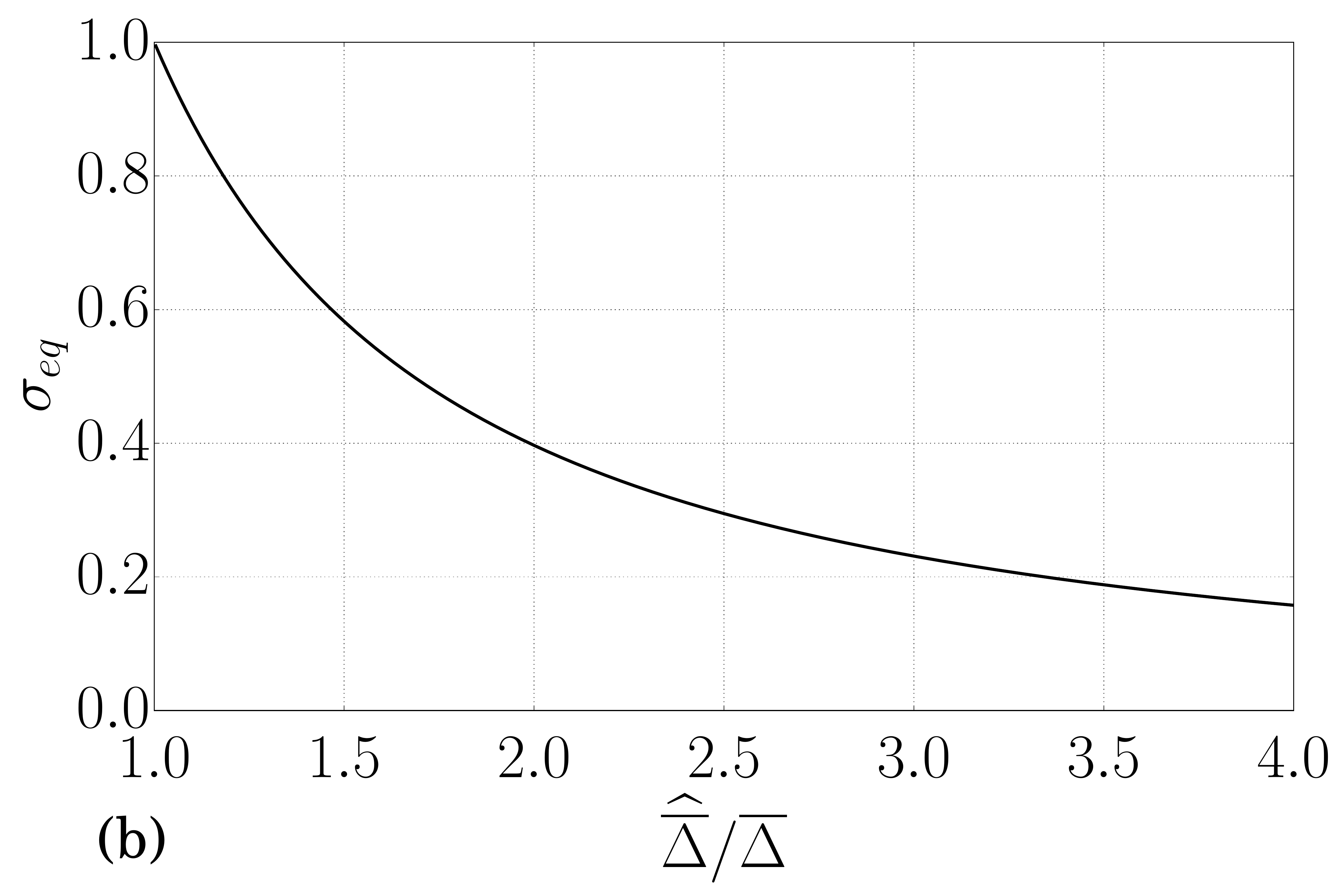}
\caption{Dependency of the sensor function $f$ on the test-filtered to grid-filtered enstrophy ratio $\sigma$ (a). Dependency of the value $\sigma_{eq}$ on the test-filter to grid-filter width ratio (b). }
\label{fsgs:f_of_sigma}
\end{figure}

The expression for the function $f(\sigma)$ is found by setting the following bounds:
\begin{itemize}
\item A lower bound for $f$ can be defined as $f(\sigma=1)=0$ where the test-filtered enstrophy is equal to the grid-filtered enstrophy.
\item The upper bound of $f$ can be calibrated by finding a value for $\sigma$ assuming a situation of fully developed isotropic turbulence, by integrating the filtered and test-filtered enstrophy in spectral space.
\end{itemize}
In the last case, the expression of the filtered enstrophy integrated over wavenumbers, for isotropic turbulence, reads:
\begin{equation}
\int_0^\infty\bar{\xi}dk=\int_0^\infty k^2E(k)\bar{G}^2(k)dk.
\end{equation}   
Substituting the expression of the Kolmogorov spectrum, the average value for $\sigma$ corresponding to an equilibrium energy cascade becomes:
\begin{equation}
\sigma_{eq}=\dfrac{\int_0^{\infty}(k\Delta)^{1/3}\hat{G}(k\Delta)\bar{G}^2(k\Delta)dk\Delta}{\int_0^\infty (k\Delta)^{1/3}\bar{G}^2(k\Delta)dk\Delta}.
\label{eq:sigeq}
\end{equation}  
The upper bound for the sensor becomes $f(\sigma=\sigma_{eq})=1$.
The following expression for $f$ is proposed:

\begin{equation}
f(\sigma) = 
\begin{cases}
1 & \mr{for}\quad\sigma<\sigma_{eq},\\
\frac{1}{2}\left(1+\sin\left(\pi\frac{\sigma_{eq}-2\sigma+1}{2(1-\sigma_{eq})}\right)\right) & \mr{for}\quad\sigma\in[\sigma_{eq},1],\\
0 & \mr{for}\quad\sigma>1.
\end{cases}
\label{eq:fsgs}
\end{equation}
The sensor function $f$ is plotted in Figure~\ref{fsgs:f_of_sigma} (left plot) showing a smooth decrease between $\sigma_{eq}$ and 1.

This shape guarantees a smooth scale-separation between large scale vortices and small-scale turbulence.
In terms of spectral content, this guarantees an increase of the eddy viscosity amplitude for higher wavenumbers.
\tblue{Such functions have been considered in the context of the development of shock sensors in Discontinuous finite element methods~\cite{persson2006sub}.}

We first derive the general expression of $\sigma_{eq}$ by developing (\ref{eq:sigeq}) assuming that the grid and test filtering operators are sharp in spectral space with respective filter lengths of $\overline{\Delta}$ and $\widehat{\overline{\Delta}}$.
This yields the following expression for $\sigma_{eq}$:
\begin{equation}
\sigma_{eq}=\cfrac{\int_0^{\pi/r_{\Delta}}(k\Delta)^{1/3}dk\Delta}{\int_0^\pi (k\Delta)^{1/3}dk\Delta}=r_{\Delta}^{-4/3},
\end{equation}  
where $r_{\Delta}=\widehat{\overline{\Delta}}/\overline{\Delta}$ is the ratio of the two filter lengths.
This result is consistent with the fact that a smaller bandwith for test-filtered scales (high values of $r_{\Delta}$) leads to an activation of the sensor for a broader range of small-scales.
This expression of $\sigma_{eq}$ is plotted in Figure~\ref{fsgs:f_of_sigma} (right plot).

In the present paper, a value of $\sigma_{eq}$ tailored for the present numerical \tblue{method} is determined, by considering the analytical expression of the test filter transfer function and the interpolation operator.
The numerical method employed (see section~\ref{sec:numerics:ns}) requires the interpolation of the vorticity components on the density locations to compute the values of enstrophy.
This interpolation introduces a damping of a part of the small-scale content of enstrophy (see the corresponding transfer function in Figure~\ref{test_filters_tf}), which can be taken into account in a modified estimate for $\sigma_{eq}$:
\begin{equation}
\label{sigmahit}
\sigma_{eq}=\cfrac{\int_0^{\infty}(k\Delta)^{1/3}\hat{G}(k\Delta)G_{\mathrm{int}}^2(k\Delta)dk\Delta}{\int_0^\infty (k\Delta)^{1/3}G_{\mathrm{int}}^2(k\Delta)dk\Delta}
\end{equation} 
where $G_{\mathrm{int}}$ is the transfer function of the interpolant.
The analytical expression of the three test filters transfer functions defined in Section~\ref{sec:testfilters} is substituted in~\ref{sigmahit} to find the corresponding values of $\sigma_{eq}$.
These values are 0.34, 0.54 and 0.71 for the GAUSS, EXPL4 and IMPL6 test fiters, respectively.

The sensor function $f$ is made spatially varying by computing $\sigma$ using a direct pointwise comparison of the filtered and test-filtered enstrophies $\hat{\bar{\xi}}(\boldsymbol{\mathrm{x}})$ and $\bar{\xi}(\boldsymbol{\mathrm{x}})$.
The sensor will therefore be active in regions where the filtered enstrophy is locally higher than the test-filtered enstrophy, identifying small-scale turbulent activity.

\subsection{Subgrid scale models} \label{sec:numerics:sgs_models}

This section summarizes the set of dissipative SGS models based on the eddy viscosity assumption that are tested with the CvP methodology.

\subsubsection{Smagorinsky} \label{sec:sgs_smag}
The most commonly employed dissipative SGS closure is the Smagorinsky model for which the eddy viscosity reads:
\begin{equation}
\mu_t=\rho(\mathrm{C_S}\overline{\Delta})^2\sqrt{\bar{\tens{S}}:\bar{\tens{S}}}
\end{equation}
where $\overline{\Delta}$ the filter width and $\mathrm{C_S}$ is the Smagorinsky parameter, which is usually found by assuming Kolmogorov turbulence equilibrium, yielding~$C_S=0.172$.

\subsubsection{Structure function} \label{sec:sgs_struct}

The second model considered is the structure function (SF) model introduced by Metais and Lesieur~\cite{metais1992spectral}, for which the eddy viscosity is given by:
\begin{equation}
\mu_t=\rho\,0.105\mathrm{C_K}^{-3/2}\overline{\Delta} \sqrt{\bar{F}_2(\overline{\Delta})}
\end{equation}
where $\mathrm{C_K}=1.5$ is the Kolmogorov constant, $\bar{F}_2(\Delta)=\left<||\bar{\vect{u}}(\vect{x})-\bar{\vect{u}}(\vect{x}+\vect{r}))||^2\right>_{||\vect{r}||=\Delta}$ the second order structure function of the velocity vector and 
$<\cdot>$ is an averaging operator involving the six neighbouring values.
The SF model is expected to be less dissipative than Smagorinsky in transitional regions and more dissipative in vortex cores.

\subsubsection{Vreman} \label{sec:sgs_vreman}

The third SGS model considered is the one by Vreman~\cite{vreman2004eddy}, which displays interesting features such as the vanishing of the subgrid dissipation in regions of pure shear and correct near wall scaling of the SGS stresses. The corresponding eddy viscosity reads:
\begin{equation}
\tred{\mu_t=\rho(2.5\mathrm{C_S})^2\sqrt{\frac{B_\beta}{\alpha_{ij}\alpha_{ij}}}}
\end{equation}
where \tred{$\alpha_{ij}=\partial u_j/\partial x_i$} is the velocity gradient tensor, \tred{$\beta_{ij}=\overline{\Delta}^2_m\alpha_{mi}\alpha_{mj}$}, $B_\beta=\beta_{11}\beta_{22}-\beta^2_{12}+\beta_{11}\beta_{33}-\beta^2_{13}+\beta_{22}\beta_{33}-\beta^2_{23}$ and $\mathrm{C_S}=0.172$, the same value as the Smagorinsky model.

\subsubsection{Dynamic model} \label{sec:sgs_dynamic}

A comparison of the CvP-LES methodology with similar state-of-the-art approach is enabled by implementing the Dynamic Smagorinsky version of Spyropoulos and Blaisdell~\cite{spyropoulos1996evaluation}.
For this model, the eddy viscosity reads:
\begin{equation}
\mu_t=\rho\mathrm{C_D}\sqrt{\bar{\tens{S}}:\bar{\tens{S}}}
\end{equation}
where $\mathrm{C_D}$ is the dynamic parameter, defined as follows~\cite{lilly1992proposed}:
\begin{equation}
\mathrm{C_D}=\frac{\left<L_{ij}M_{ij}\right>}{\left<M_{kl}M_{kl}\right>}
\end{equation}
where $<\cdot>$ is a spatial averaging operator acting over directions of statistical homogeneity and $L_{ij}$, $M_{ij}$ are tensors defined as follows:
\begin{equation}
L_{ij}=\tf{\bar{\rho}\tilde{u}_i\tilde{u}_j}-\frac{1}{\hat{\bar{\rho}}}\tf{\bar{\rho}\tilde{u}_i}\tf{\bar{\rho}\tilde{u}_j}
\end{equation}
\begin{equation}
M_{ij}=-2 \tf{\f{\Delta}}^2 \hat{\bar{\rho}} \left|\hat{\bar{S}}\right| \left(\hat{\bar{S}}_{ij}-\frac{1}{3}\hat{\bar{S}}_{kk}\right) + 2\f{\Delta}^2 \reallywidehat{ \bar{\rho} \left|\bar{S}\right| \left(\bar{S}_{ij}-\frac{1}{3}\bar{S}_{kk}\right) }
\end{equation}
The test filter used to computed $M_{ij}$ and $L_{ij}$ is the IMPL6 compact filter described in section~\ref{sec:numerics:filters}.
\tred{We note that 21 test-filtering operations are required in order to compute all terms appearing in the tensors $M_{ij}$ and $L_{ij}$.}

%% file: results.tex

\clearpage
\newpage

\section{CvP-LES of Taylor-Green vortex breakdown} \label{sec:results_TGV}

\subsection{Problem definition}
The Taylor-Green vortex flow features the breakdown of large-scale vortices into broadband small-scale turbulence.
This test case is therefore of interest to assess the ability of subgrid models to both characterize accurately transition and fully developed turbulence.
The initial conditions are defined in a cubic domain $\Omega=[-\pi L,\pi L]^{3}$ as: 

\begin{equation}
\mathbf{u}(\mathbf{x},0)=\begin{pmatrix}V_{0}\sin\left(x/L\right)\cos\left(y/L\right)\cos\left(z/L\right)\\
-V_{0}\cos\left(x/L\right)\sin\left(y/L\right)\cos\left(z/L\right)\\
0
\end{pmatrix},
\end{equation}
\begin{equation}
\mathrm{p}\left(\mathbf{x},0\right)=\mathrm{p_{0}}+\frac{\rho_{0}V_{0}^{2}}{16}\left[\cos(2x/L)+\cos(2y/L)\right]\left(\cos(2z/L)+2\right).
\end{equation}
\begin{equation}
\mathrm{\rho}\left(\mathbf{x},0\right)=\rho_0.
\end{equation}

Computations are carried out in dimensionless form by using the reference length $L$, velocity magnitude $V_{0}$ and density $\rho_{0}$.
The Reynolds number $Re=\rho_{0}V_{0}L/\mu_{0}$ is set to 5000 and the Mach number $M=V_{0}/\sqrt{\gamma\,p_0/\rho_0}$ is set to 0.1, to avoid compressibility effects.
The state of turbulence is monitored by evaluating the temporal evolution of the volume-averaged total kinetic energy $E$ and the dissipation rate $\varepsilon$:
\begin{equation}
E(t)=\frac{1}{2|\Omega|}\int_\Omega\vect{u}(\vect{x},t)\cdot\vect{u}(\vect{x},t)d\vect{x},
\end{equation}
\begin{equation}
\varepsilon(t)=\frac{dE(t)}{dt}.
\end{equation}
The temporal evolution of spatially integrated subgrid dissipation $\varepsilon_{\mathrm{SGS}}$ is considered as well to monitor the intensity of the modeled subgrid dissipation:
\begin{equation}
\varepsilon_{\mathrm{SGS}}(t)=\frac{1}{|\Omega|}\int_\Omega\mu_t\tens{S}:\tens{S}d\vect{x}.
\end{equation}

\subsection{Evaluation of the CvP-LES accuracy}
In this section, the accuracy of the CvP approach coupled with the Smagorinsky model is evaluated in comparison with the Dynamic model.
The DNS performed by Chapelier and Lodato~\cite{chapelier2016spectral} featuring $480^3$ degrees of freedom is chosen as the reference computation.
Three LES computations featuring $72^3$ grid points are performed with the baseline Smagorinsky model, the CvP approach coupled with Smagorinsky and the Dynamic model.
A simulation is also carried out without SGS closure on the same grid, termed as no model computation.
It is notable that the LES resolution is more than six times coarser in each direction than the reference DNS calculation.

\begin{figure}[!h]
\centering
\includegraphics[width=\linewidth]{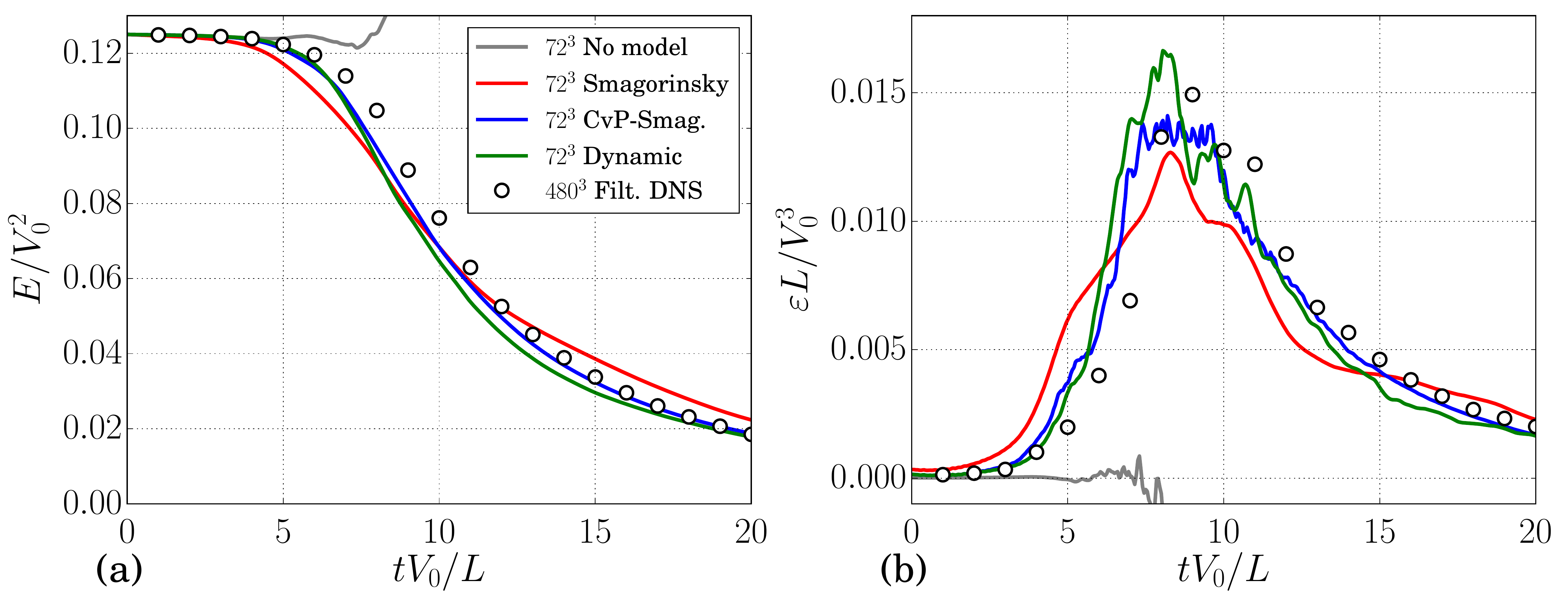}
\caption{Evolution of the turbulent kinetic energy (a) and dissipation (b) for the LES of the Taylor-Green vortex.}
\label{TGV_stats}
\end{figure}

\begin{figure}[!h]
\centering
\includegraphics[width=\linewidth]{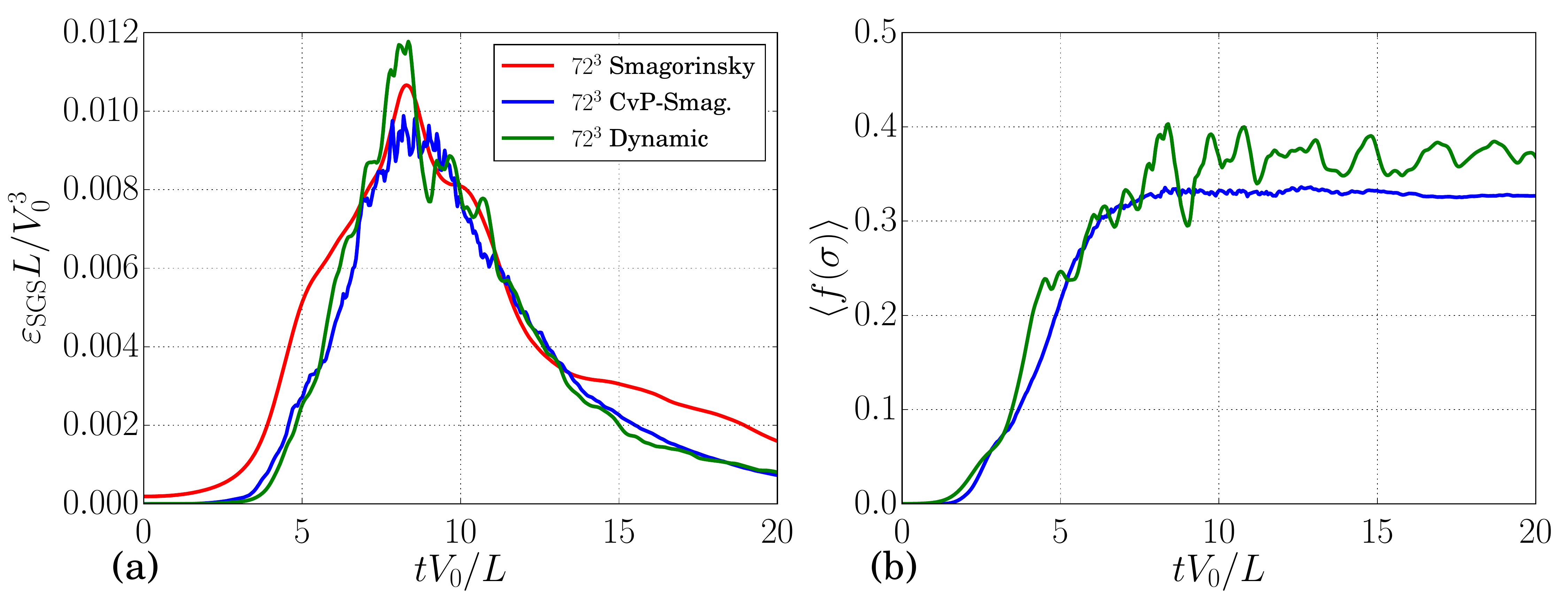}
\caption{Temporal evolution of the subgrid dissipation (a) and the dynamic parameter defined as the factor of the classical Smagorinsky eddy viscosity ($\left<f\right>$ for the CvP sensor and $\mathrm{C_D}/(\mathrm{C_S}\Delta)^2$ for the Dynamic model)(b) for the LES of the Taylor-Green vortex flow. }
\label{TGV_sensor}
\end{figure}

Figure~\ref{TGV_stats} depicts the evolution of the volume averaged turbulent kinetic energy (or TKE) and dissipation for the three LES computations and the DNS.
The DNS data is filtered using a spectral sharp cutoff filter with width matching the LES grid size.
The DNS filtered dissipation is obtained by the means of the temporal derivative of the filtered DNS energy.
The computation performed without SGS closure leads to a blow-up of the TKE and negative values of dissipation.
\tblue{The Smagorinsky computation overestimates the dissipation at early stages, and underestimates it afterwards. The early excess in dissipation is due to an immediate activation of the subgrid dissipation as seen in Figure~\ref{TGV_sensor}, when the flow is still dominated by large-scale dynamics. This in turn impairs the development of the small-scale content of the flow, leading to an under-estimation of the peak of dissipation which depends strongly on the accurate representation of small-scale dynamics.}
The Smagorinsky model coupled with the CvP approach successfully improves the quality of the prediction for both \tblue{the kinetic energy and dissipation}.
Figure~\ref{TGV_sensor} presents the temporal evolution of the mean values of the subgrid dissipation and the CvP sensor function $f(\sigma)$.
The subgrid dissipation for the sensor and dynamic models is efficiently lowered at the beginning of the computation as opposed to the Smagorinsky computation.
\tblue{Small-amplitude oscillations in the subgrid dissipation of the CvP-Smagorinsky computation are observed near the peak of dissipation. This behavior could be related to the spatial variations of the CvP sensor as opposed to the volume-averaged strategy required for the computation of the dynamic parameter of the Germano model. However, the oscillations seem to occur only during the onset of the small scales, as the subgrid and total dissipations for the CvP-Smagorinsky case show a smooth evolution at the early and late stages of the computation.}
It is found that $f(\sigma)$ is low at the beginning of the computation, showing that the sensor successfully detects the presence of large coherent vortices and mitigates the subgrid dissipation accordingly. 
\tblue{By comparing figures~\ref{TGV_stats}b and~\ref{TGV_sensor}a, it is noteworthy that the subgrid dissipation contributes to most of the total dissipation for this particular discretization and Reynolds number. This shows that the amplitude of the subgrid stresses is higher than the viscous stresses and confirms that the present case is fit to assess the performance of SGS models.}

\begin{figure}[!h]
\centering
\includegraphics[width=\linewidth]{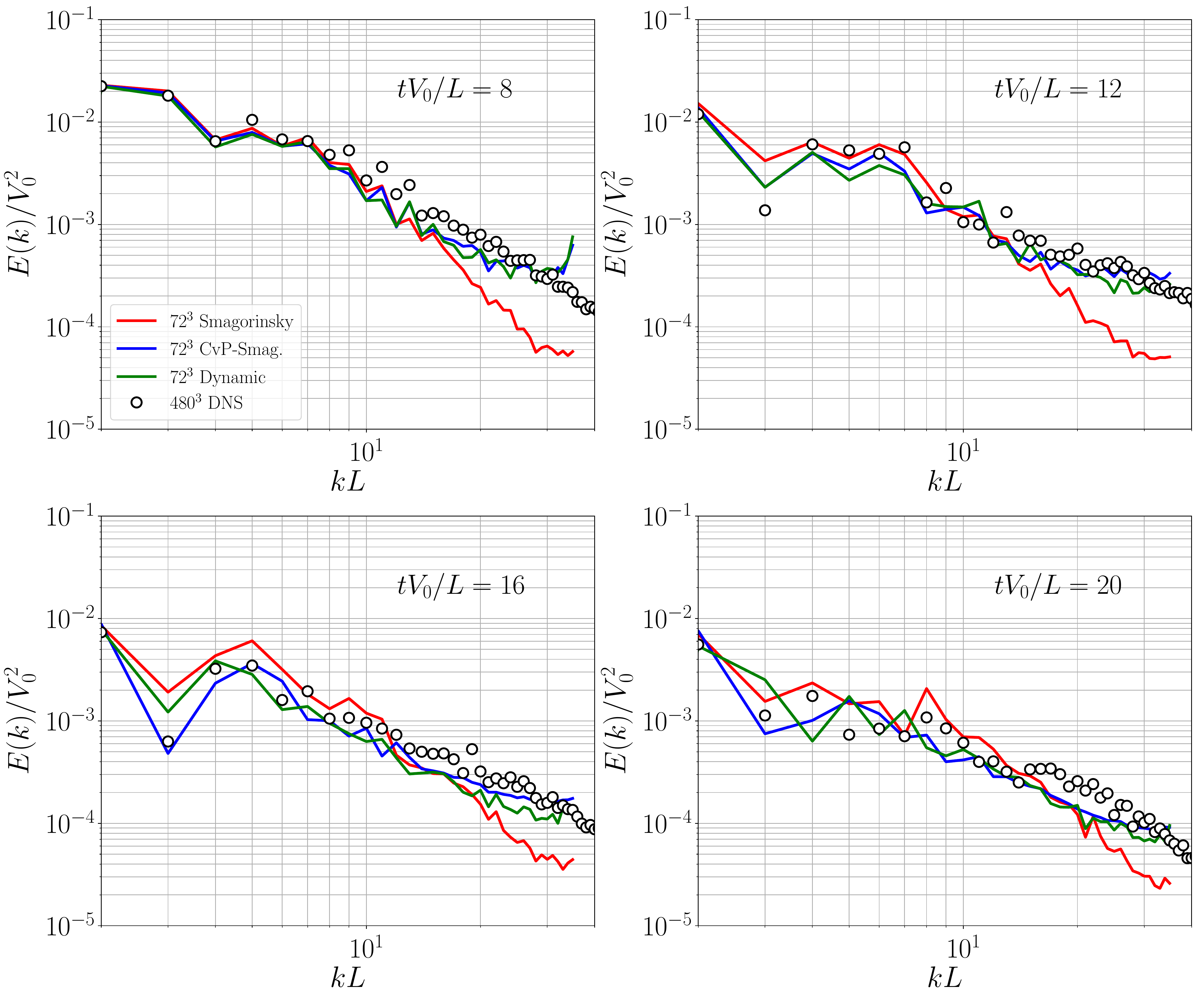}
\caption{Energy spectra at various times for the LES computations and reference DNS of the Taylor-Green vortex at $Re=5000$. LES computations are performed using a $72^3$ grid.}
\label{spc_tgv}
\end{figure}

The energy spectra computed at different times ranging from $tV_0/L=8$ to $tV_0/L=20$ are plotted in Figure~\ref{spc_tgv}.
At $tV_0/L=8$, near the peak of enstrophy, while all models provide a good prediction of the large-scale dynamics, the small-scale turbulent activity is significantly damped in the Smagorinsky computation.
The Dynamic and CvP computations improve greatly the behavior of the Smagorinsky model and show a good match with the DNS at the small-scale level.
The CvP-LES approach is slightly more accurate than the Dynamic model in the midrange of the energy spectrum.
The two models also exhibit a slight high-wavenumber pile-up of energy.
This phenomenon due to aliasing errors rapidly vanishes as the flow evolves towards fully developed turbulence.
At later times, the Smagorinsky model shows an imbalance of spectral energy distribution, overestimating and underestimating the large and small-scale energy content, respectively.
The strong damping of small-scale energy is likely to create a bottleneck effect that blocks the energy transfers, responsible for the observed spectral energy imbalance.
The energy spectra at later times are more accurately predicted by the Dynamic and CvP models, and the CvP-LES, in particular, correctly preserves the large-scale energy for long-time integration.

\subsection{Computational cost analysis}
In this section, the computational costs of the different SGS modeling approaches adopted in the previous section are assessed.
All calculations on the $72^3$ grid where performed in parallel using 16 computational cores.
The no-model CPU time is chosen as the reference as it is the faster run. 
Table~\ref{cputime} reports the relative CPU times for the no-model, Smagorinsky, CvP-Smagorinsky and Dynamic models runs.
As expected, a computational overhead is observed when SGS models are active.
Remarkably, the CvP-Smagorinsky yields a computational overhead comparable to the one observed for the baseline Smagorinsky model.
This shows that the computation of the sensor function $f(\sigma)$ is inexpensive, which is expected as the only costly step is the test filtering of the enstrophy field.
The Dynamic model, comparing similarly in terms of accuracy, is found to be more expensive, leading to a computational overhead of 40$\%$.
This excessive cost is due to the test filtering needed to compute the tensors $L_{ij}$ and $M_{ij}$ as well as averaging globally the numerator and denominator of the dynamic coefficient expression.

\begin{table}[t]
\centering
\begin{tabular}{c|cccc}
Model & CPU Time & Computational overhead \tabularnewline
\hline 
\hline 
No SGS model & $t_{ref}$ & -  \tabularnewline
Smagorinsky &  $1.134t_{ref}$ & +13.4\% \tabularnewline
CvP+Smagorinsky & $1.151t_{ref}$ & +15.1\% \tabularnewline
Dynamic model & $1.404t_{ref}$ & +40.4\%  \tabularnewline
\end{tabular}
\caption{Computational time for the various modeling approaches considered for the Taylor-Green vortex computations using a $72^3$ grid.}
\label{cputime}
\end{table}

\subsection{Test-filter width sensitivity study}
\begin{figure}[!h]
\centering
\includegraphics[width=\linewidth]{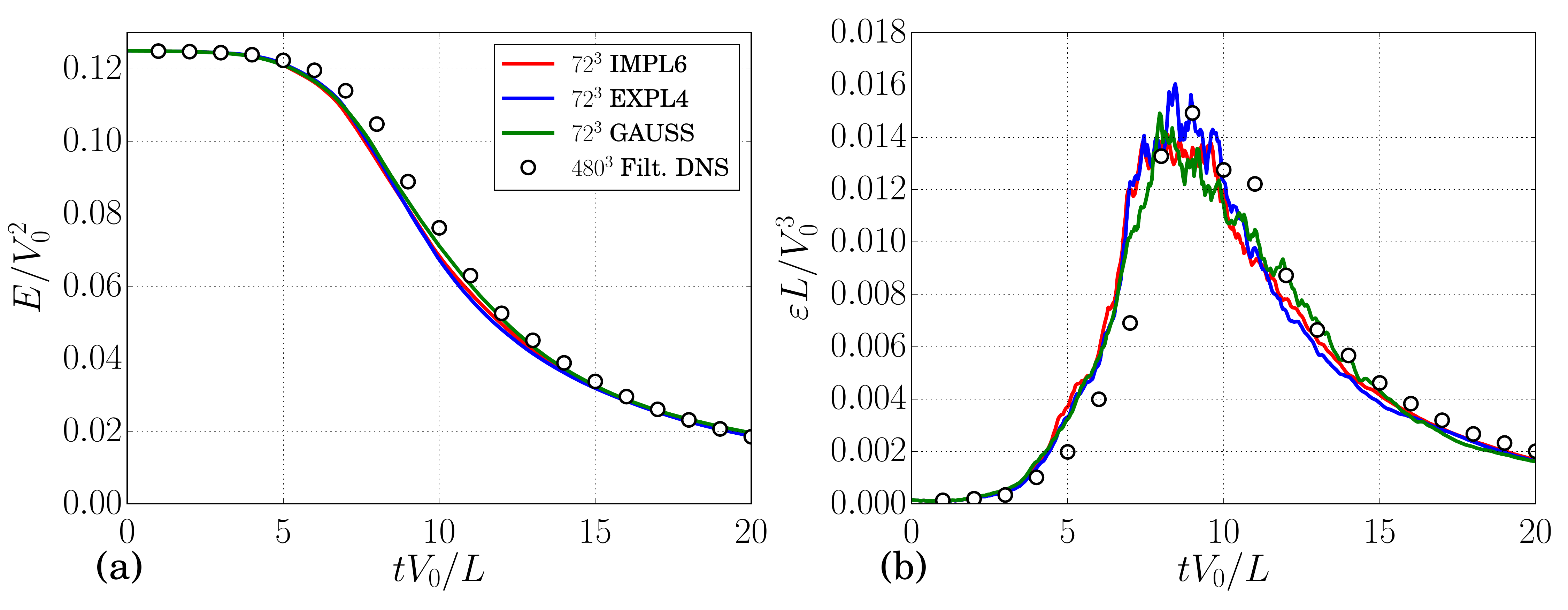}
\includegraphics[width=0.5\linewidth]{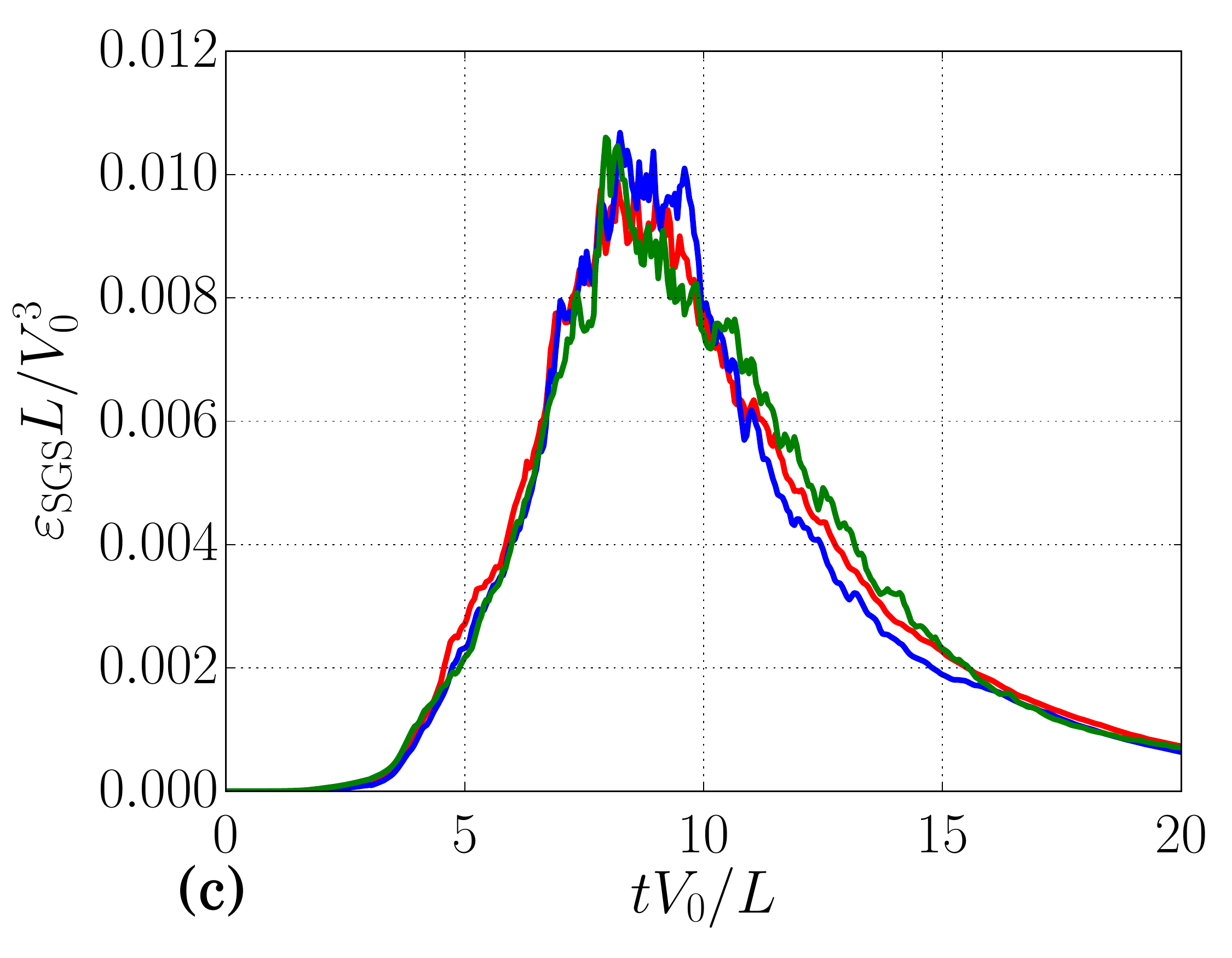}
\caption{Evolution of the kinetic energy (a), kinetic energy dissipation (b) and subgrid dissipation (c) for the LES of the Taylor-Green vortex at $Re=5000$ using the CvP-Smagorinsky model with different test filters.}
\label{tgv_testfilt}
\end{figure}

In this section, the three test filters defined in section~\ref{sec:testfilters} are considered for the LES of the Taylor-Green vortex flow.
All CvP-LES computations shown here are performed on a $72^3$ grid.
Figure~\ref{tgv_testfilt} shows the evolution of the dissipation for these computations.
The choice of the test filter has a marginal impact on the quality of the solution.
The robustness of the CvP methodology to the test-filter width is very encouraging as it makes its successful extension to other numerical schemes very likely.

\subsection{Influence of the subgrid model}
\begin{figure}[!h]
\centering
\includegraphics[width=\linewidth]{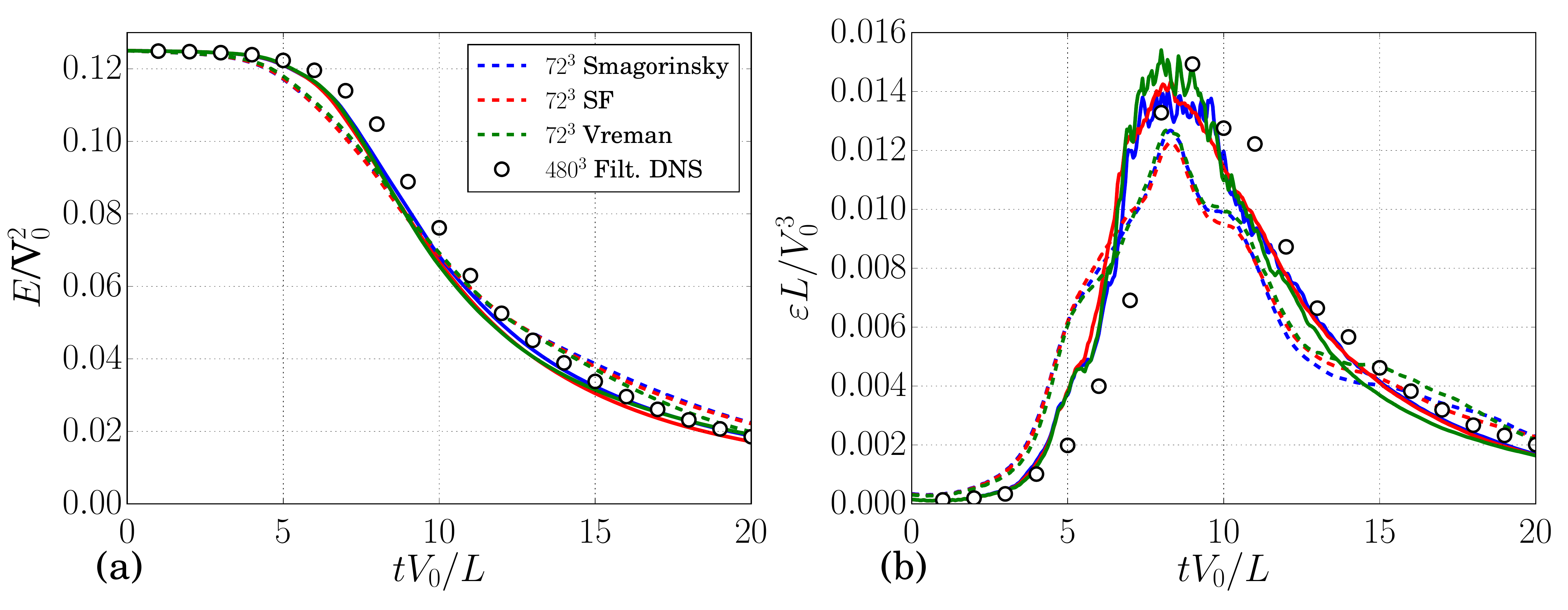}
\caption{Mean kinetic energy (a) and dissipation (b) for CvP (solid lines) and non-CvP (dashed lines) LES computations of the Taylor-Green vortex at $Re=5000$.}
\label{TGV_comp_mod}
\end{figure}

\begin{figure}[!h]
\centering
\includegraphics[width=\linewidth]{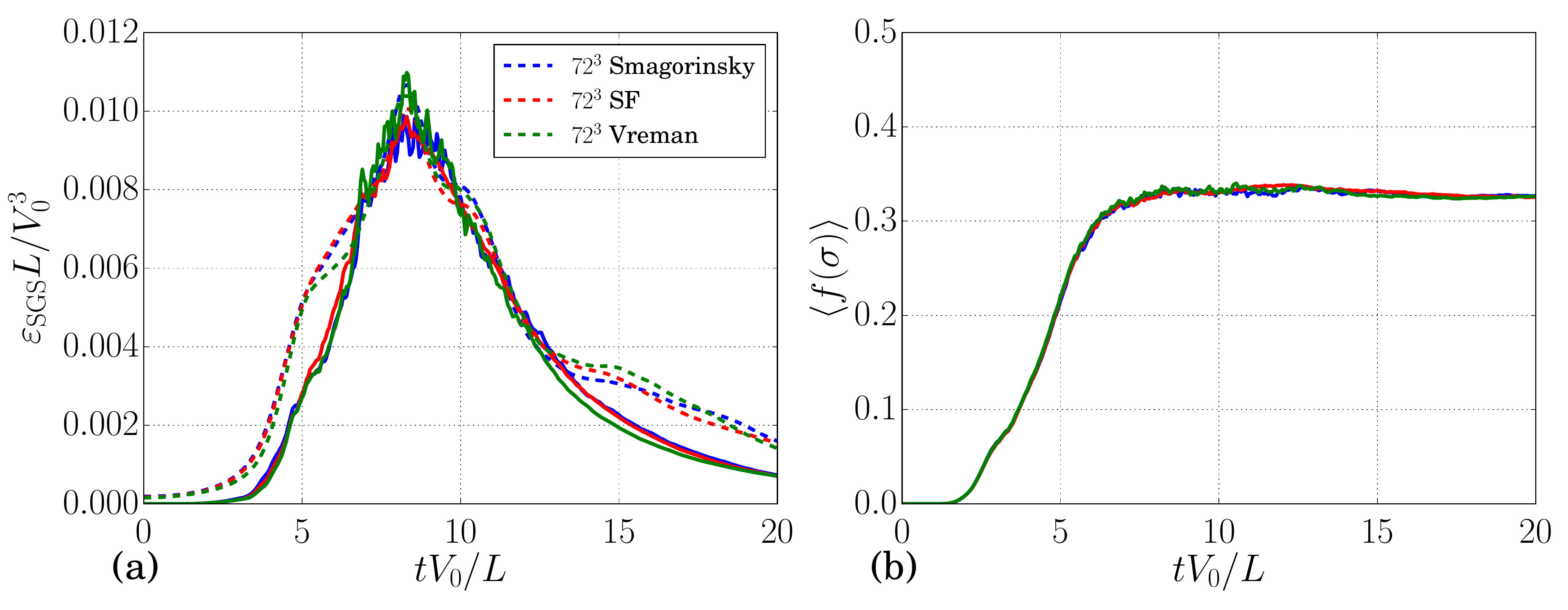}
\caption{Temporal evolution of the subgrid dissipation (a) and CvP sensor function (b) for the LES of the Taylor-Green vortex flow.}
\label{TGV_sensor_mod}
\end{figure}

In the present section, the ability of the CvP methodology to enhance the performance of the subgrid models described in section \ref{sec:sgs_smag} is assessed.
Fig.~\ref{TGV_comp_mod} presents the evolution of the TKE and dissipation LES computations with and without CvP.
The baseline Smagorinsky and Structure function models lead to similar results, which is expected for freely decaying turbulence.
Both models however introduce a strong SGS dissipation which leads to an erroneous prediction of dissipation at the early stages of the computations, emphasizing their inaptitude to capture the transient features of the flow (characterized by a model subgrid dissipation active during the transition).
For all models considered, the CvP sensor yields a clear improvement in the prediction of the total dissipation rate due to a reduction of the SGS dissipation at the early stages, see Figure~\ref{TGV_sensor_mod}.

\subsection{Grid sensitivity study}
\begin{figure}[!h]
\centering
\includegraphics[width=\linewidth]{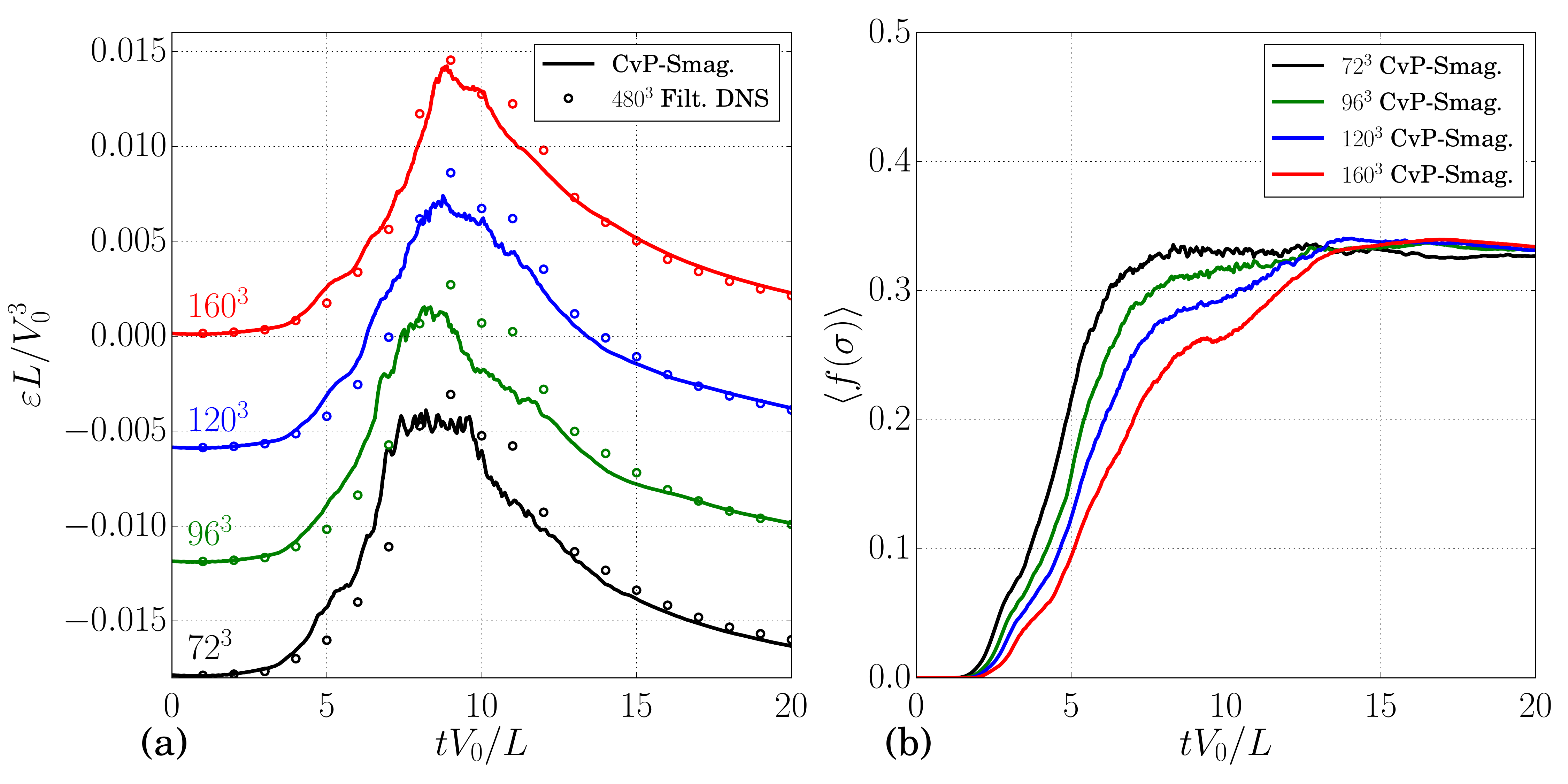}
\caption{Evolution of the dissipation (a) and the CvP sensor function (b) for the mesh convergence study of the Taylor-Green vortex LES at $Re=5000$. \tblue{The dissipation values for the $72^3$, $96^3$ and $120^3$ computations are vertically shifted for clarity (resp. -0.018, -0.012 and -0.006).}}
\label{mesh_conv}
\end{figure}

The accuracy of the newly developed approach is here assessed on different grid sizes.
Additional CvP-Smagorinsky computations are performed using \tblue{$96^3$, $120^3$ and $160^3$ grid points.}
The results are compared to DNS data with matching grid-filter widths.
Figure~\ref{mesh_conv} presents the evolution of the dissipation for the CvP-LES and filtered DNS.
An excellent agreement with the filtered DNS is found for all \tblue{four meshes considered, showing the ability of the present approach to yield consistent results when the grid-filter cutoff wavenumber is located in different regions of the inertial range.}
Right plot of Figure~\ref{mesh_conv} also shows the evolution of the sensor function for all \tblue{four} meshes.
The sensor function is activated at later times for the finer meshes, which is consistent with the notion that an increase in resolution can sustain smaller scales, delaying the development of all the relevant scales in the flow.
The sensor function reaches a plateau value independent of the resolution for the decaying regime, which means that fully developed turbulence is correctly detected for the \tblue{four} meshes.

\clearpage
\newpage

\section{CvP-LES of Double helical vortex breakdown} \label{sec:results:helical_vortex}

\begin{figure}[h!]
\centering
\includegraphics[width=0.49\linewidth]{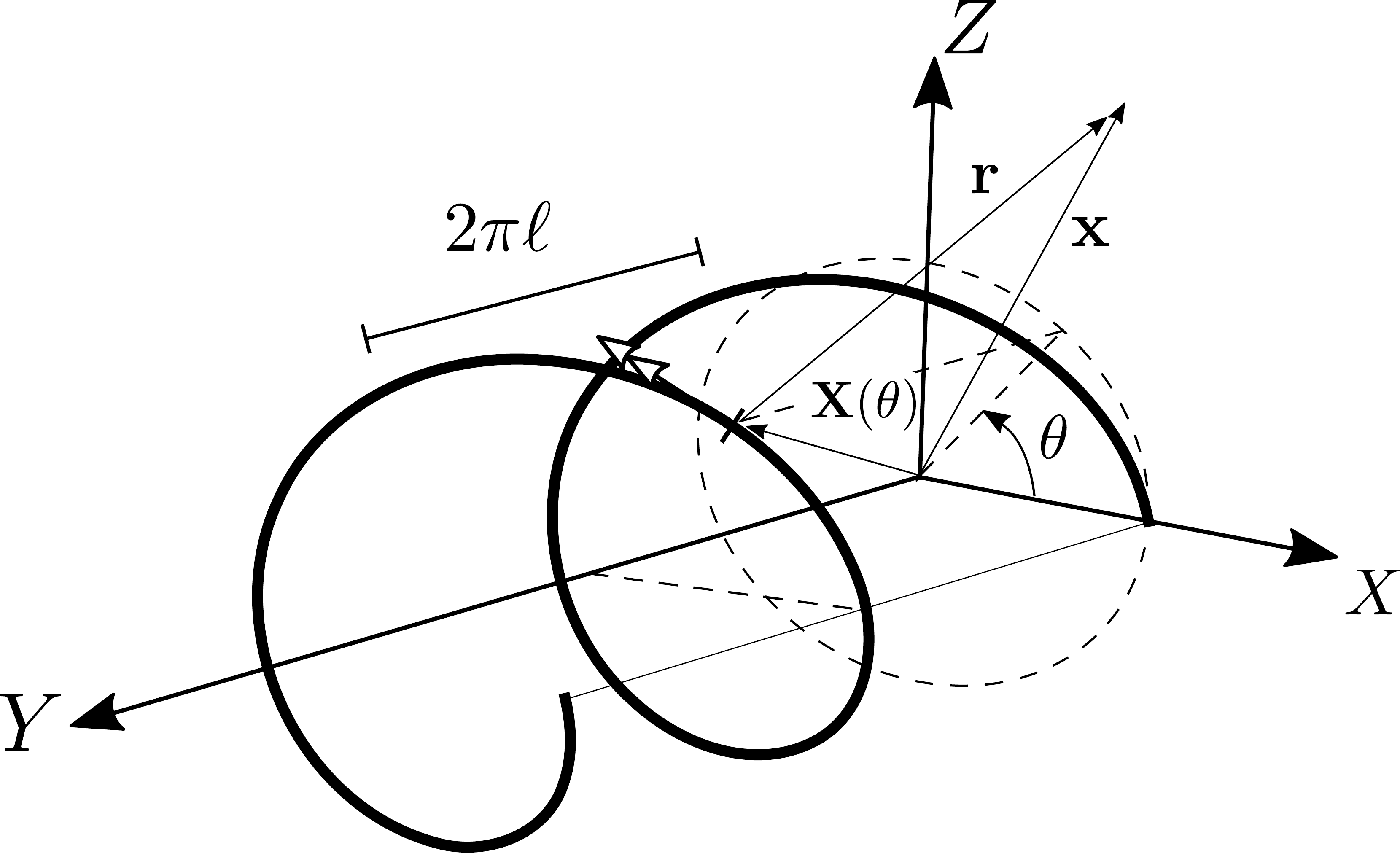}
\includegraphics[width=0.49\linewidth]{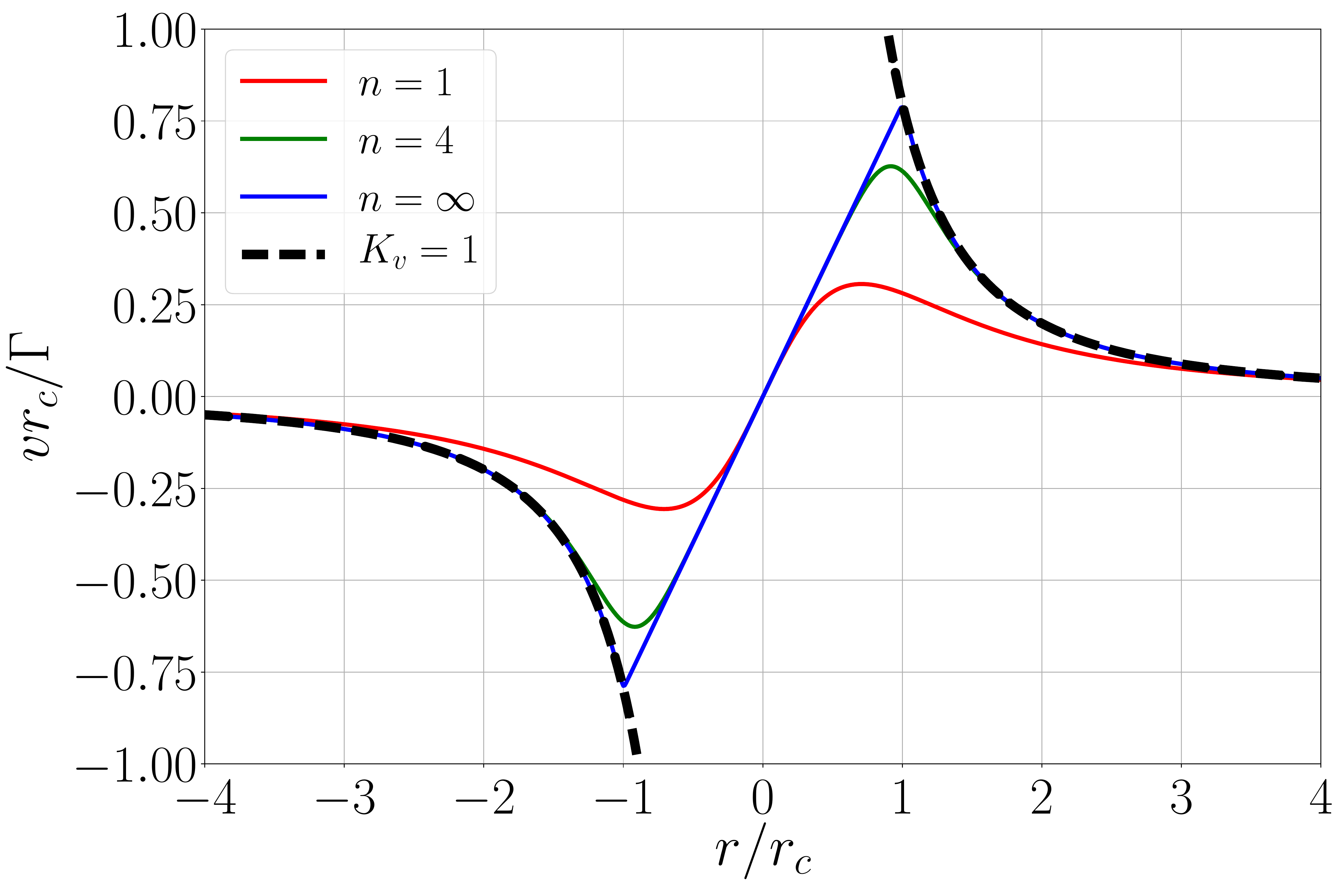}
\caption{Helical vortex setup (left), velocity profile as a function of radius for the Biot-Savart law modified with a kernel function to account for the viscous core (right).}
\label{vort_setp}
\end{figure}

\subsection{Problem definition}
In this section, a fundamental test case is defined for the study of helical vortices that are representative of the wake of rotating devices.
A vortex filament is initialized in a triply periodic box.
The parametric curve describing the vortex filament reads:
\begin{equation}
\bold{X}(\theta) = \left[ R\cos(\theta),\ell \theta,R\sin(\theta)\right]^{\mathrm{T}}
\end{equation}
where $R$ is the radius and $h=2\pi l$ is the pitch of the helix.
The velocity field induced by the vortex filament is determined by the Biot-Savart law:
\begin{equation} \label{eq:biot_savart}
\bold{u}(\bold{x})=-\frac{\Gamma}{4\pi}\int{K_v\frac{\left(\bold{x}-\bold{X}(\theta)\right)\times\bold{t}(\theta)}{\left|\bold{x}-\bold{X}(\theta)\right|^3}d\theta},
\end{equation}
where $\bold{t}(\theta)=(-R\sin{\theta},\ell,R\cos{\theta})$ is the tangent vector to the helical filament, $\Gamma$ is the circulation and $K_v$ is a smoothing kernel defining the shape of the vortex core~\cite{bagai1993flow} which reads: 
\begin{equation} \label{eq:smoothing_kernel}
K_v=\frac{\left|\bold{x}-\bold{X}(\theta)\right|^3}{\left(\left|\bold{x}-\bold{X}(\theta)\right|^{2n}+r_c^{2n}\right)^{\frac{3}{2n}}},
\end{equation}
where $r_c$ is the core radius.
The case $n=\infty$ corresponds to a Rankine vortex.
The value of $n=4$  is adopted to achieve a smooth transition between the inner, rotational flow and the outer, potential flow.
The corresponding plots of transverse velocity $v$ as a function of the radial distance from the vortex core are shown in Figure~\ref{vort_setp} for different values of $n$ and for the case $K_v=1$ which corresponds to the Biot-Savart law without correction, leading to infinite velocity at the core of the vortex. 

\subsection{LES simulations}

In this section, the experimental measurements of the double helical vortex instability studied by Nemes et al.~\cite{nemes2015mutual} are reproduced using the presently developed numerical framework.
The radius of the helix is $R=0.115$m, as in the experiment.
The ratio of the helical pitch to helix radius is set to $h/R=1.1$ which guarantees an unstable configuration due to the mutual inductance phenomenon.
The ratio of the vortex core radius to helix radius is set to $r_c/R=0.06$ to match the experimental conditions.
Finally, the Reynolds number based on the circulation is set to $Re_{\Gamma}=\Gamma/\nu=7000$.
A direct numerical simulation of this flow is unfeasible due to the Reynolds considered.
LES are performed using the sensor approach coupled to the Smagorinsky model using discretizations with \tblue{$128^3$, $192^3$ and $256^3$ grid points and a cubic box of dimensions $0.5^3$, resulting in respectively 3.5, 5 and 7 grid points inside the vortex cores.}

\begin{figure}[!h]
\centering
\includegraphics[width=0.8\linewidth]{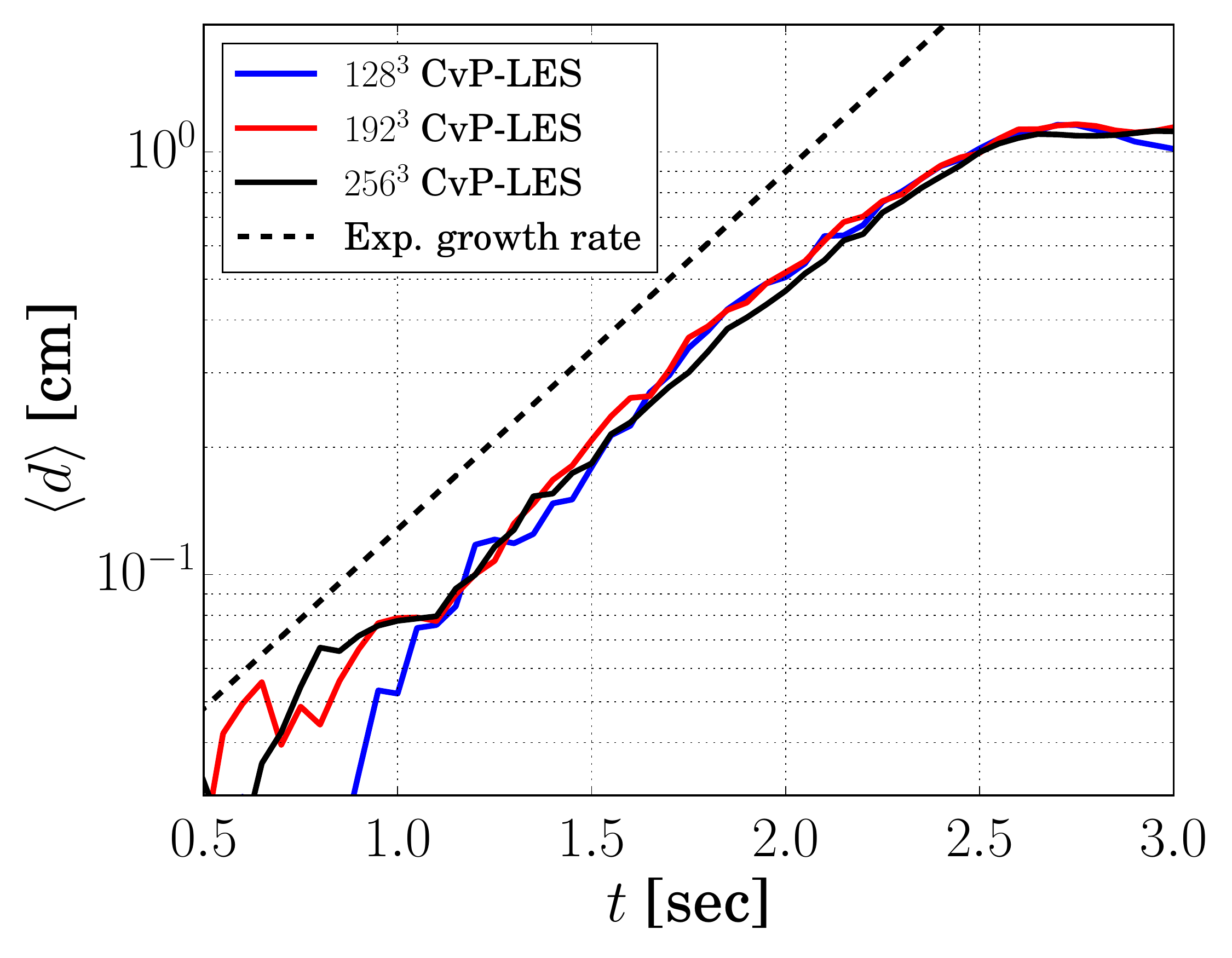}
\caption{Deviation of vortex core position from their initial position as a function of time. CvP-LES of the double helical vortex configuration. \tblue{The dashed exponential evolution corresponds to the experimentally observed deviation rate by Nemes et al.~\cite{nemes2015mutual}}} 
\label{growth_rate}
\end{figure}

To quantitatively compare the experiment to the present LES computation, the growth rate of the instability is calculated from the deviation of the position of vortex cores compared to their initial position, as proposed by Quaranta et al.~\cite{quaranta2015long}. 
\tblue{The deviation of the vortex cores is measured as:
\begin{equation}
d(t)=\frac{1}{L_Y}\int_0^{L_Y}|r(y,t)-R|dy
\end{equation}
where $r(y,t)$ is the position of maximal vorticity magnitude in planes $(x,z)$.} 
Figure~\ref{growth_rate} presents the evolution of the vortex deviation measured at different times.
The data is plotted along the experimental growth rate measured by Nemes et al.~\cite{nemes2015mutual}.
For all resolutions considered, the LES is found to match correctly the experimental growth rate between $t=0.5s$ and $t=2.5s$.
Oscillations are observed for the coarser discretizations for which the precise evaluation of the vortex core is difficult due to the reduced number of grid points.
After $t=3s$, the flow enters the non-linear regime corresponding to the development of smaller scales around the main helices.

\begin{figure}[!h]
\centering
\includegraphics[width=0.37\linewidth]{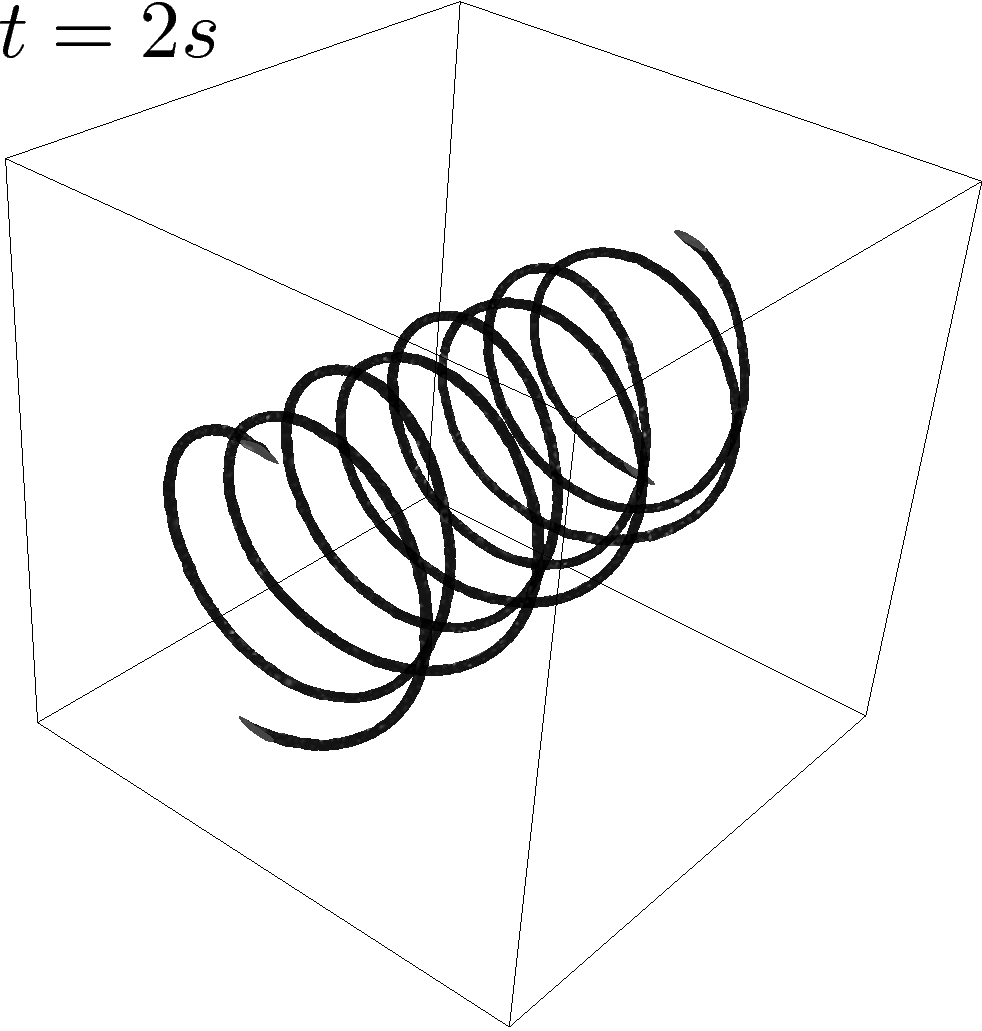}
\includegraphics[width=0.37\linewidth]{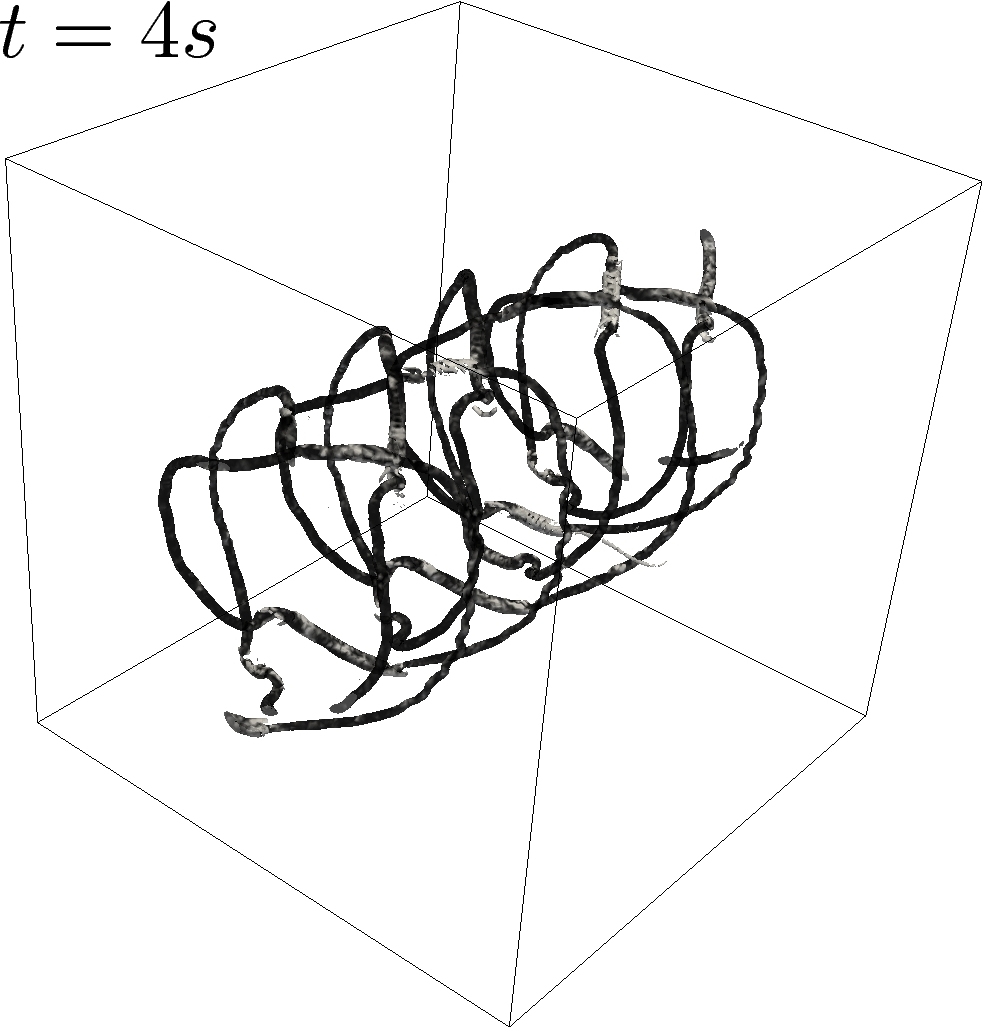}
\includegraphics[width=0.37\linewidth]{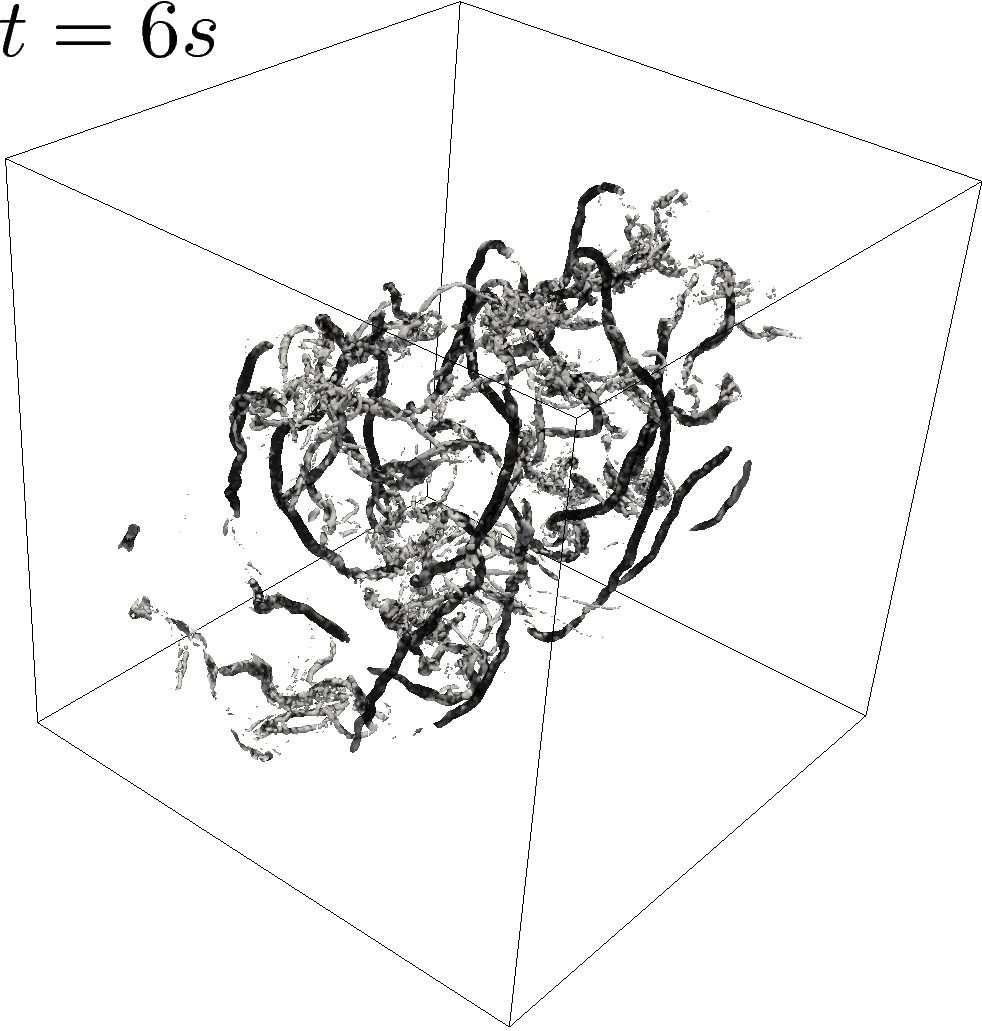}
\includegraphics[width=0.37\linewidth]{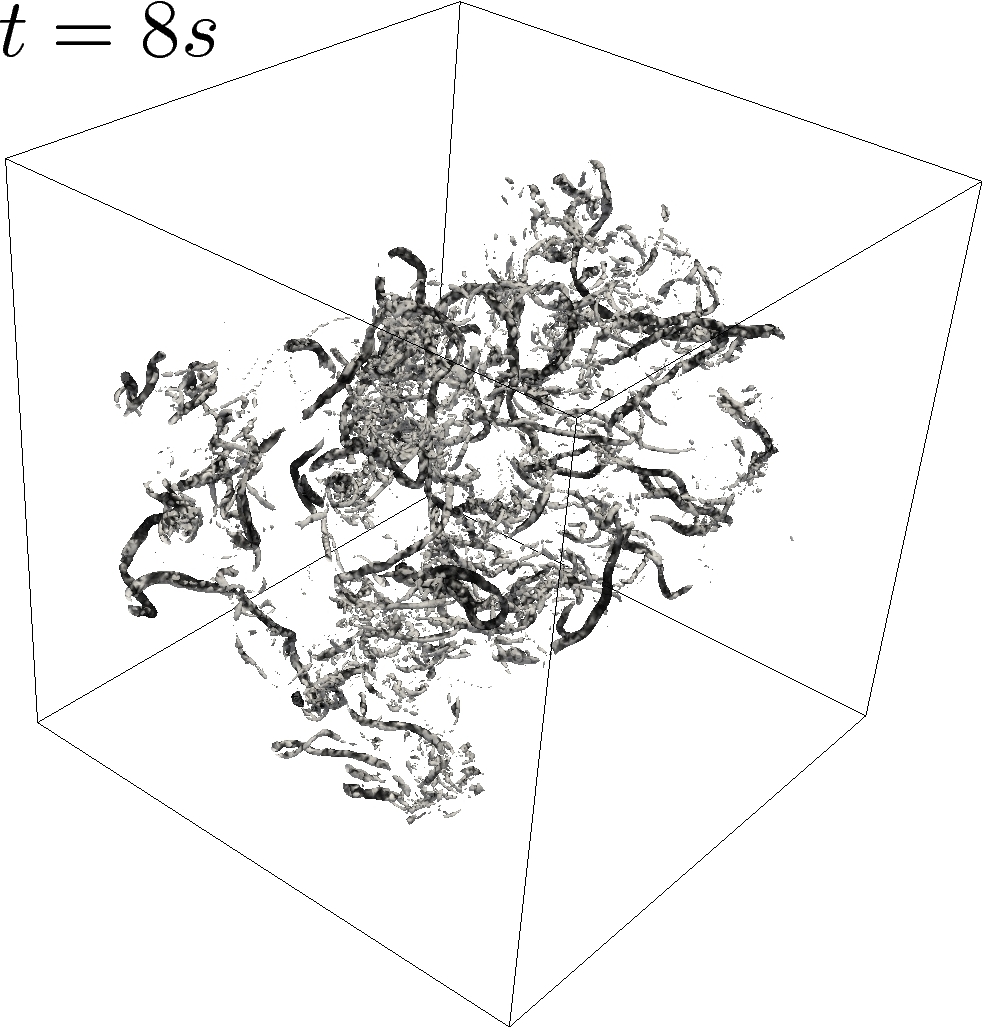}
\includegraphics[width=0.37\linewidth]{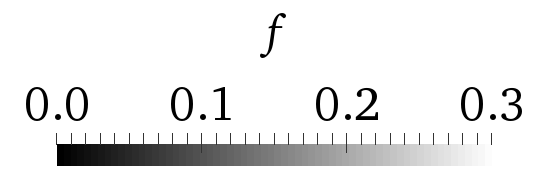}
\caption{Iso-surfaces of vorticity colored by the CvP sensor function at different times for the LES of the double helix configuration, \tblue{$192^3$ discretization}.}
\label{helix_sensor}
\end{figure}

\begin{figure}[!h]
\centering
\includegraphics[width=0.3\linewidth]{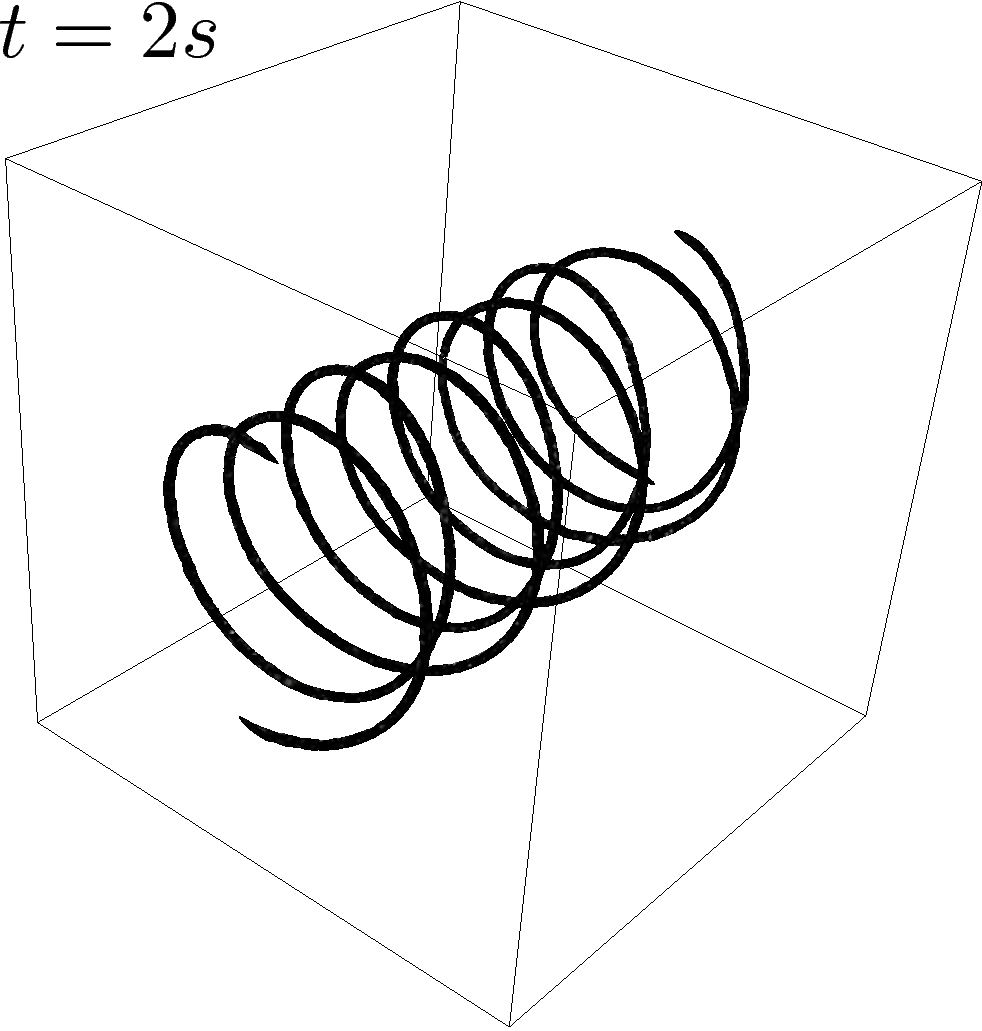}
\includegraphics[width=0.3\linewidth]{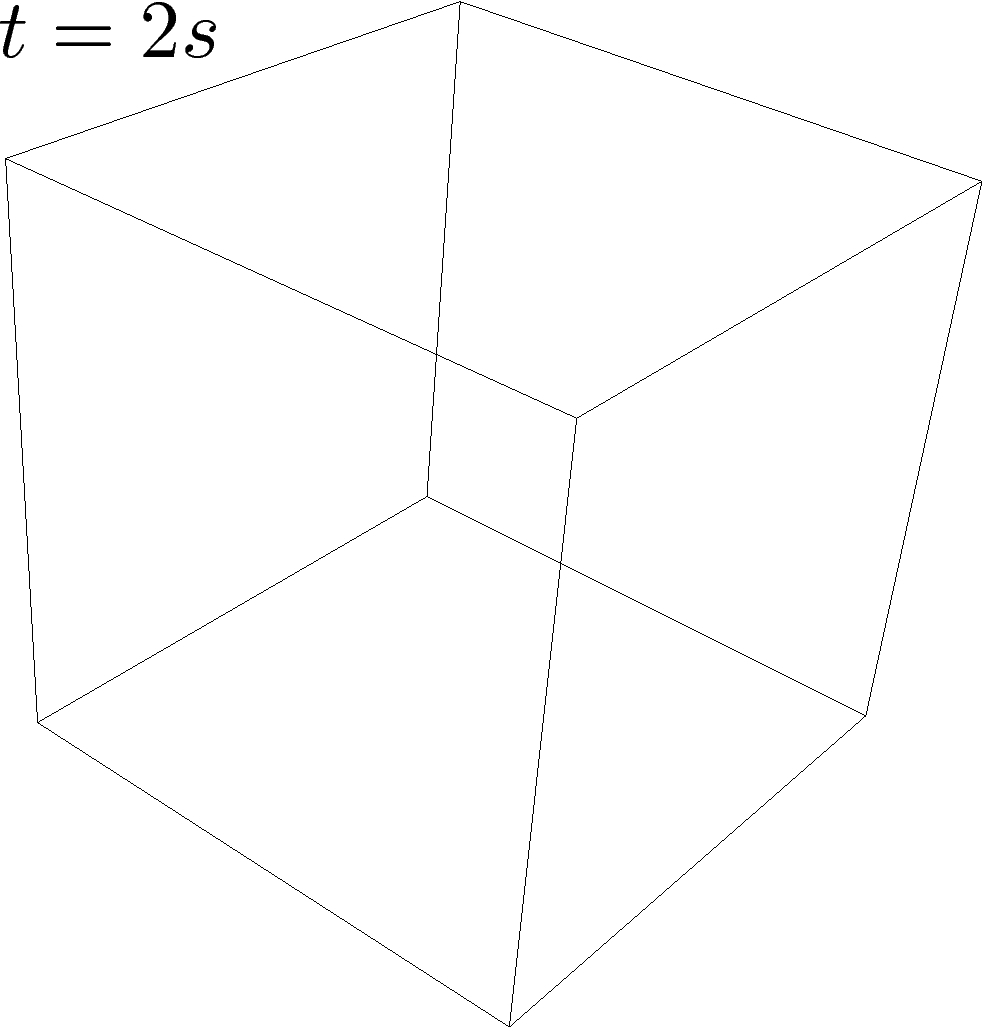} \\
\includegraphics[width=0.3\linewidth]{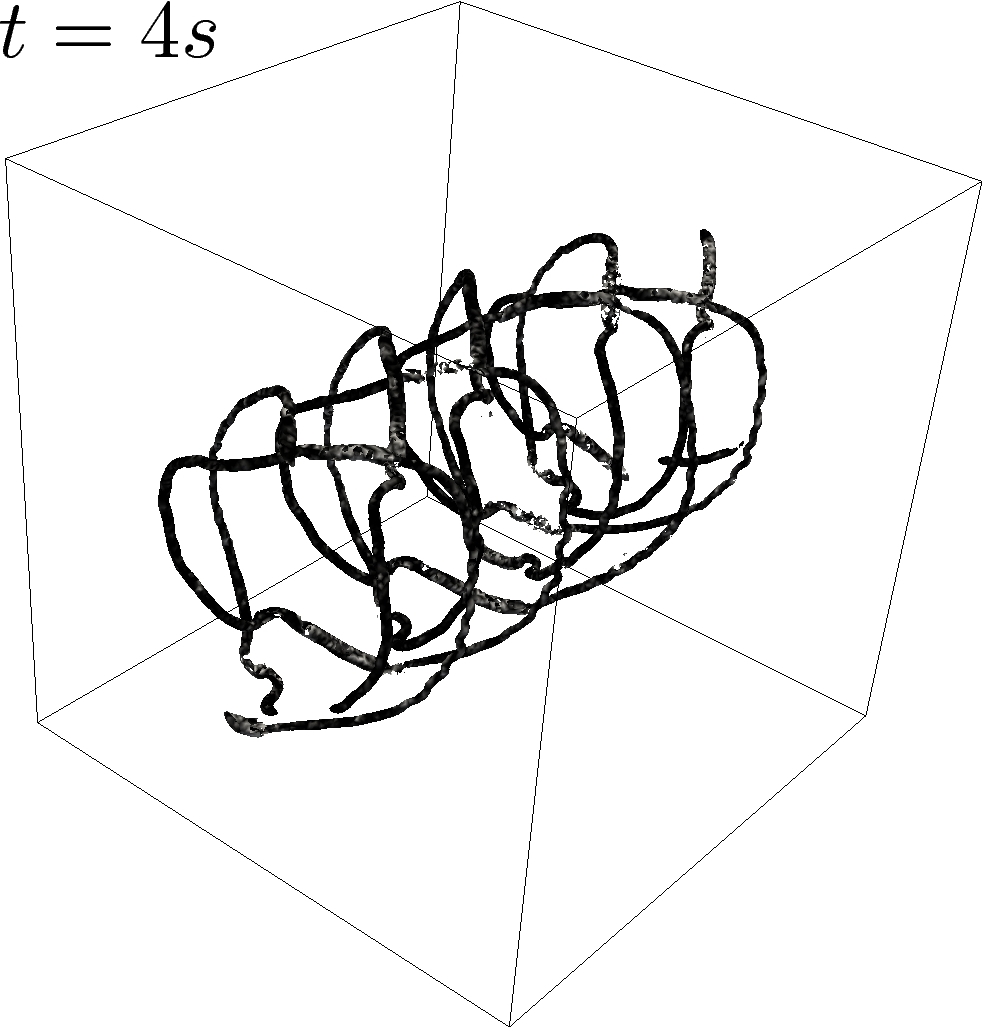}
\includegraphics[width=0.3\linewidth]{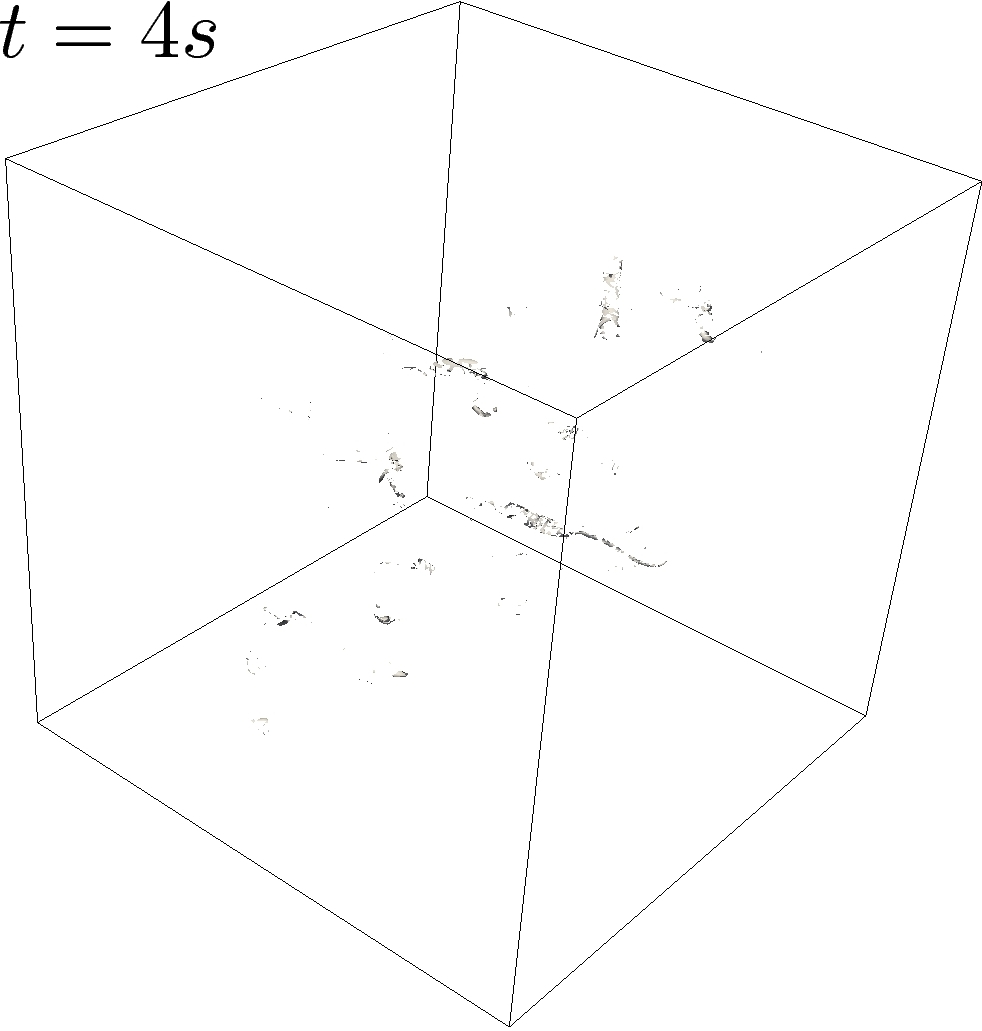} \\
\includegraphics[width=0.3\linewidth]{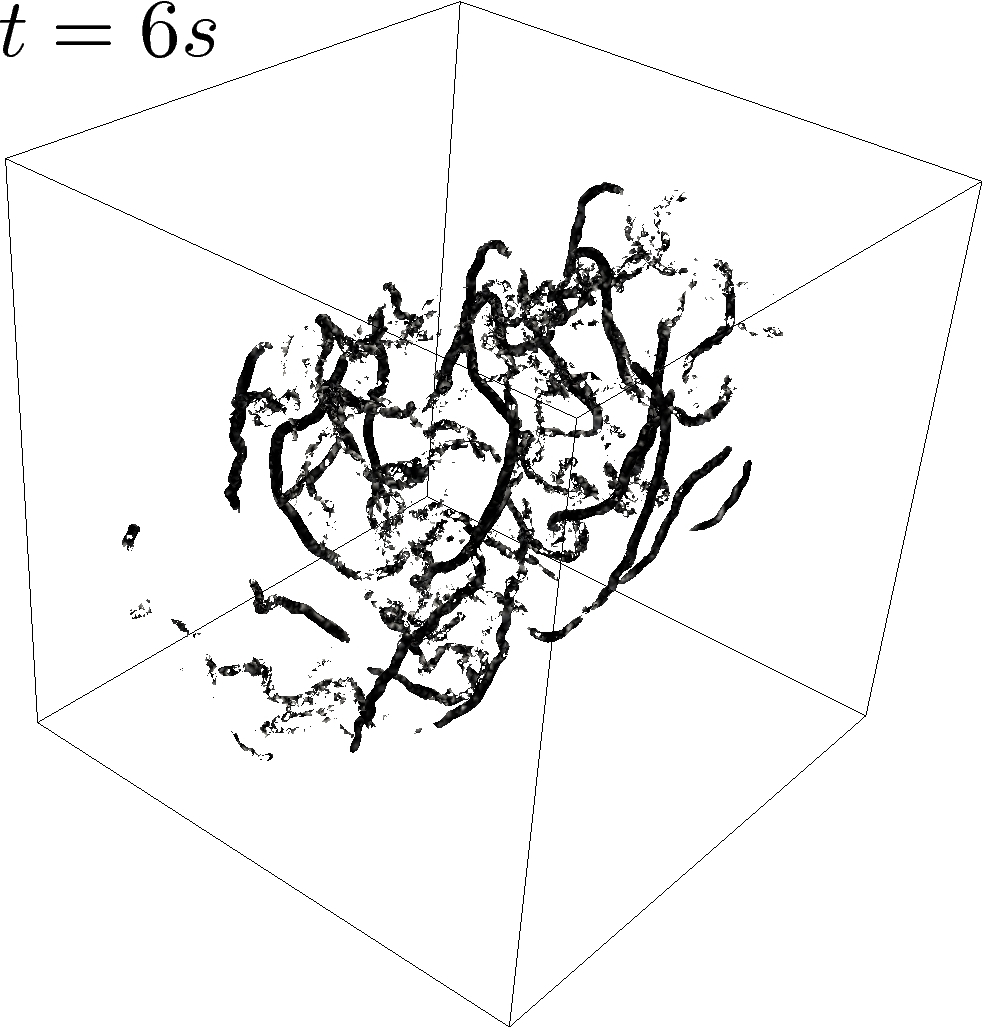}
\includegraphics[width=0.3\linewidth]{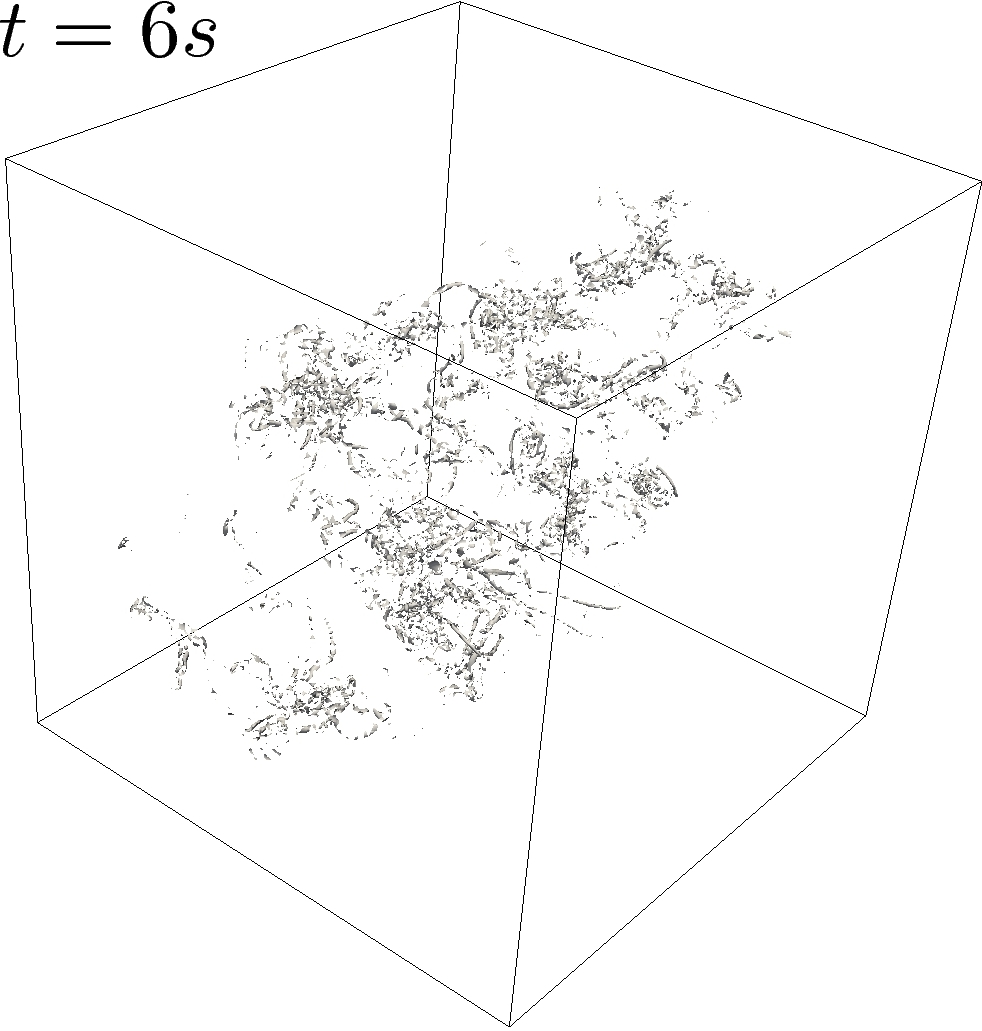} \\
\includegraphics[width=0.3\linewidth]{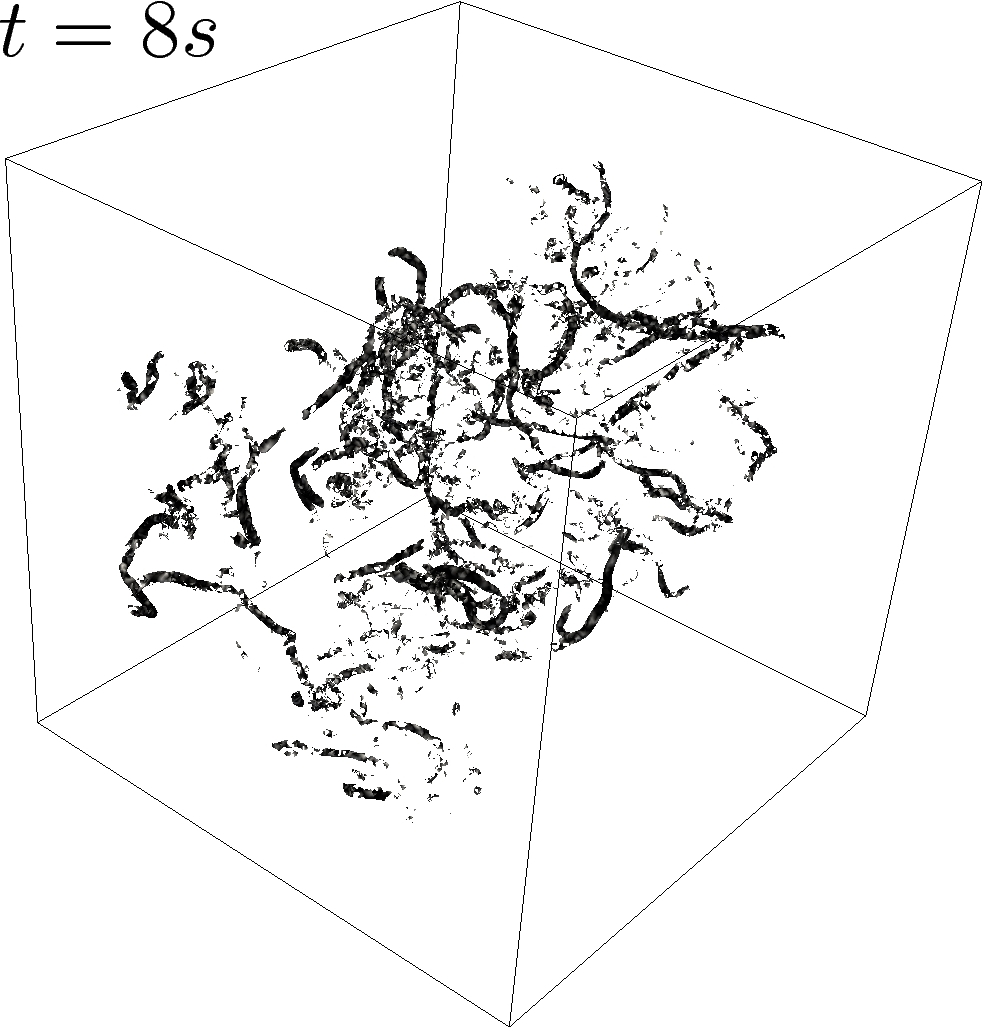}
\includegraphics[width=0.3\linewidth]{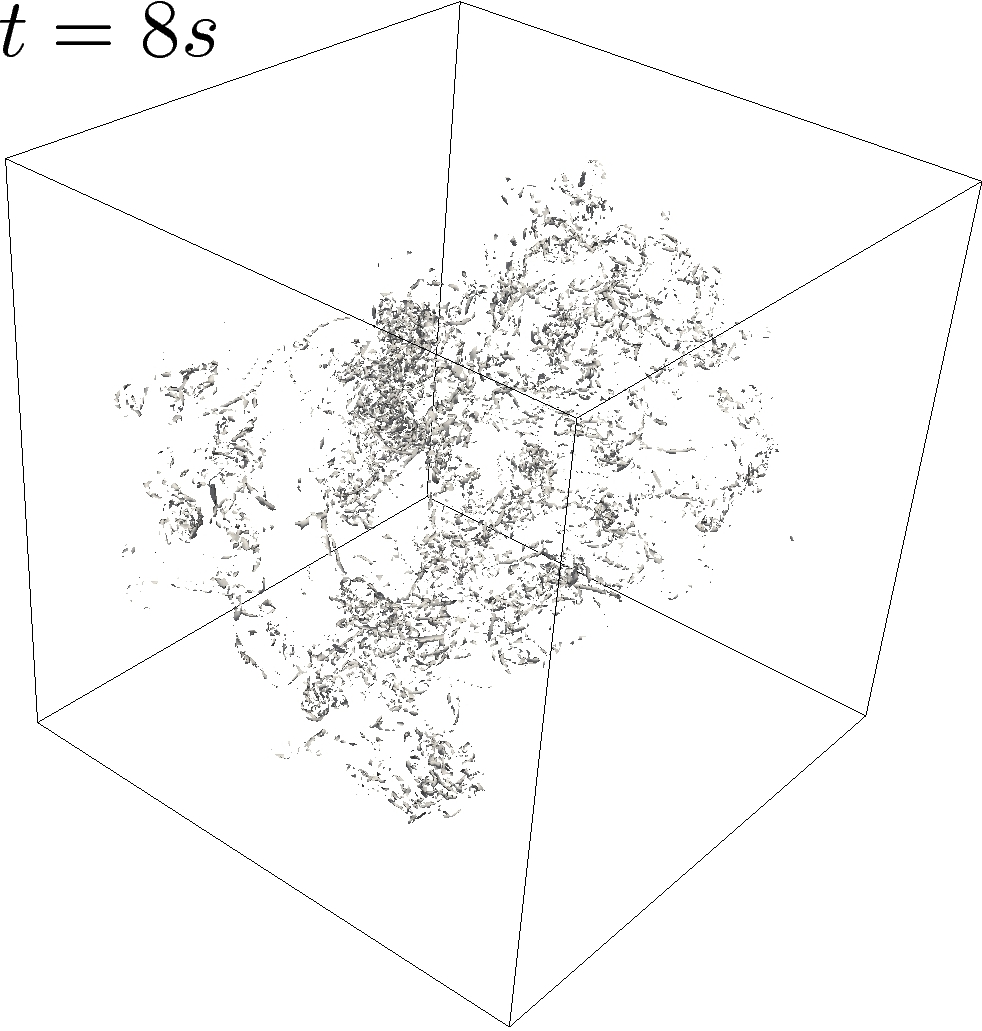} \\
\caption{\tblue{Identification of the large and small turbulent scales using the CvP sensor} from the LES of the double helix configuration, \tblue{$192^3$ discretization}. Left column: iso-surfaces of vorticity \tblue{clipped} to values of $f(\sigma)$ between 0 and 0.3. Right column: iso-surfaces \tblue{clipped to values higher than 0.3.}}
\label{helix_sensor_clip}
\end{figure}

The capacity of the sensor function to identify the large scale motion of the flow is assessed by inspecting vorticity iso-surfaces colored by the sensor function $f(\sigma)$ from the $128^3$ computation.
Figure~\ref{helix_sensor} presents the iso-surfaces at four different times.
At the beginning of the computation, only large scale features are present in the flow.
The onset of small-scales due to the instability caused by mutual inductance is visible at the time $t=6s$.
For the early stages of the computation, the iso-surfaces are mainly colored in black, meaning that the sensor function $f(\sigma)$ is mostly zero.
When the flow transitions and small-scales develop, high-values of the sensor function (white regions) start to appear.
Remarkably, the CvP sensor is able to separate the small-scales from the coherent motion at all times.
This feature is also visible at $t=8s$, when the large scales begin to vanish and the small-scale motion becomes prominent.
Furthers visualizations in Figure~\ref{helix_sensor_clip} show the iso-surface of vorticity clipped, corresponding respectively to low and high values of the sensor function.
The tracking of coherent motion through time is well identified by low values of $f$ as, on the other hand, high values of $f$ monitor the onset of small scales as time evolves and as coherent vortices breakdown to fully developed turbulence.
The small-scales finally become dominant is the flow, which is characterized by high values of the sensor function in a broad region of the computational box.

\clearpage
\newpage

\section{\tblue{CvP-LES of wall-bounded turbulence}} \label{sec:results:channel}

\subsection{Description of the test case}
This section features the application of the CvP methodology to wall-bounded turbulence.
A compressible turbulent channel flow case is considered.
The computational domain is $\Omega=[0,L_X]\times[0,2h]\times[0,L_Z]$ where $h$ is the channel's half width.
For the present calculations, the values $L_X=14h$ and $L_Z=8h$ are chosen.
Periodic conditions are considered in the longitudinal and transverse directions, and the top and bottom walls are no-slip isothermal. The flow conditions are the ones of Coleman {\it et al.}~\cite{coleman1995numerical}, with bulk Reynolds number $Re_b=\rho_bU_bh/\mu_w=3000$ and bulk Mach number $M_b=U_b/\sqrt{\gamma r T_w}=1.5$. The resulting friction Reynolds number is $Re_\tau=\rho_w u_\tau h/\mu_w=220$.
The mass flow rate is kept constant by imposing an artificial pressure gradient as a source term in the longitudinal momentum equation.

\subsection{\tblue{A priori evaluation and correction of the CvP sensor for wall-bounded turbulence}}

\begin{figure}
\centering
\includegraphics[width=0.98\linewidth]{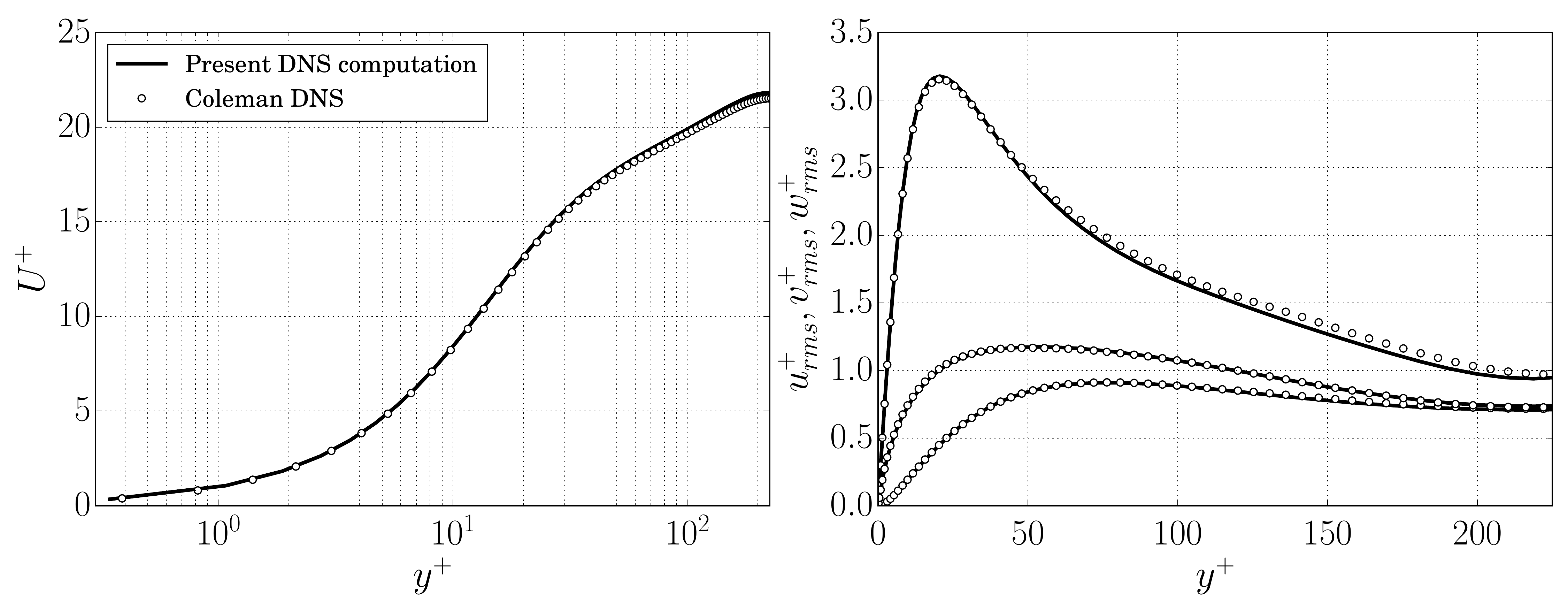}
\caption{\tblue{Mean and fluctuating velocity profiles for the reference DNS of the compressible channel flow at $Re_b=3000$ and $M_b=1.5$.}}
\label{channel_dns_valid}
\end{figure}

\begin{figure}
\centering
\includegraphics[width=0.7\linewidth]{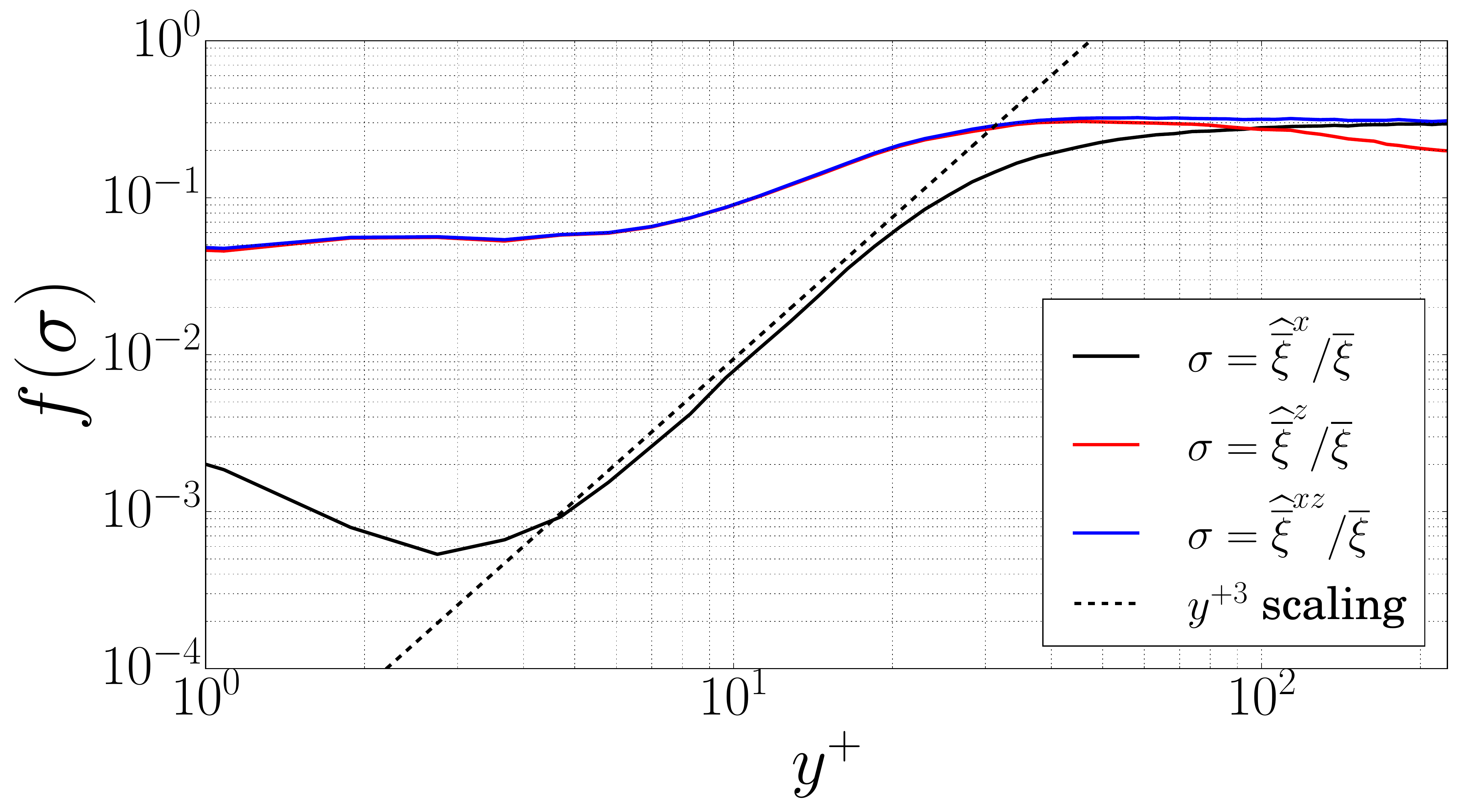}
\caption{\tblue{CvP sensor function computed a priori from the DNS of the compressible channel flow at $Re_b=3000$ and $M_b=1.5$.}} 
\label{channel_cvp_function}
\end{figure}

\tblue{In this section, the performance of the CvP sensor for wall-bounded turbulence is assessed via an a priori evaluation of the function $f(\sigma)$ from a DNS computation at the flow conditions described in sectio 6.1. The DNS computation features a $128\times 96 \times 128$ grid yielding the following grid spacings in wall units: $\Delta_x^+=24$, $\Delta_z^+=14$ and $\Delta y_{\mathrm{min}}^+=0.35$. This resolution is comparable to the one considered by Coleman et al., which is: $\Delta_x^+=19$, $\Delta_z^+=12$ and $\Delta y_{\mathrm{min}}^+=0.1$. Figure~\ref{channel_dns_valid} shows a validation of the present DNS against the results of Coleman et al.~\cite{coleman1995numerical}. The agreement is satisfactory; the slight discrepancies observed are due to the differences in numerical methods and grid resolutions considered.
The plane-averaged value of CvP sensor function $f(\sigma)$ extracted from this newly created DNS dataset is plotted against $y^+$ in Figure~\ref{channel_cvp_function} considering anisotropic variants to the definition of $\sigma$: to evaluate the sensor's behavior in inhomogeneous flow, we consider two other formulations of the sensor built on the directional filtering of $\sigma$. The two variants consist in evaluting $\sigma$ by either test filtering the resolved enstrophy $\overline{\xi}$ in the streamwise or spanwise direction only, resulting in respectively $\widehat{\overline{\xi}}^{x}$ and $\widehat{\overline{\xi}}^{z}$. It is seen that computing $\sigma$ as $\sigma_z=\widehat{\overline{\xi}}^{z}/\overline{\xi}$ yields high values of $f$ in the vicinity of the wall, which means that the near-wall high-wavenumber enstrophy is high. However, computing $\sigma$ as $\sigma_x=\widehat{\overline{\xi}}^{x}/\overline{\xi}$, yields a correct $y^{+^3}$ scaling for the sensor function near the wall, except for $y^+<4$. The usual formulation for which the enstrophy is test-filtered in both directions yields the same near-wall behavior as the spanwise-only test-filtered version. A generic way to improve the near-wall scaling for the CvP-LES approach could be devised from the previous observations, by removing the test-filtering in the direction carrying the highest small-scale enstrophy. This will be the topic of future investigation. In the next section, the CvP-Smagorinsky approach with $\sigma$ built from the test-filtering in the streamwise direction only is assessed for a LES computation of the channel.}
\tblue{A straightforward approach for applying the CvP method to wall-bounded turbulence is to consider the usual construction of the CvP sensor with test-filtering in all directions and subgrid models that provide a vanishing subgrid dissipation near-wall, such as the Vreman model. In this case, an improvement of the subgrid model can still be expected as the CvP sensor function amplitude is reduced near the wall. This approach is assessed as well in the next section.}



\subsection{LES of the compressible channel flow}
\begin{figure}
\centering
\includegraphics[width=0.98\linewidth]{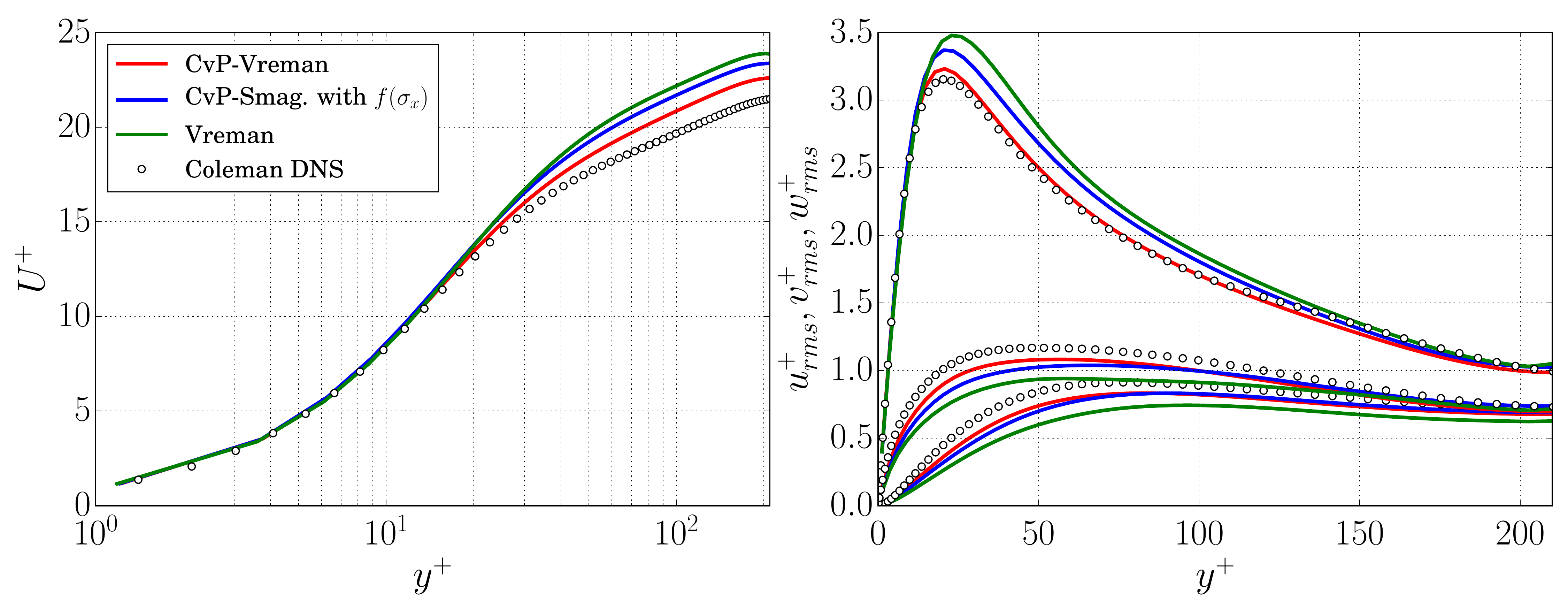}
\caption{Mean and fluctuating velocity profiles for the LES of the compressible channel flow at $Re_b=3000$ and $M_b=1.5$.} 
\label{channel}
\end{figure}
\tblue{In this section, the CvP methodology is evaluated for the LES of the channel flow using the Vreman and Smagorinsky models, with a model constant set to a value $C_\mathrm{S}=0.172$, which is the usual value considered for isotropic turbulence or free shear flows. For the CvP-Vreman computation, $f(\sigma)$ is built from the test-filtered enstropy in all directions, while for the CvP-Smagorinsky computation, $\sigma$ is buit from the test-filtered enstrophy in the streamwise direction only.} 
\tblue{The LES resolution is coarsened about two times in each direction compared to the DNS and resulting in grid spacings of $\Delta_x^+=48$, $\Delta_z^+=28$ and $\Delta y_{\mathrm{min}}^+=1.3$. Three computations are performed, namely a LES of the channel flow using the Vreman model, a LES using the CvP-Vreman approach and a LES using the CvP-Smagorinsky approach.}
\tblue{Figure~\ref{channel} presents the mean and fluctuating velocity profiles for the LES computations compared to the reference DNS. The profiles are made non-dimensional by dividing the velocity by the friction velocity $u_\tau=\sqrt{\tau_w/\rho_w}$.}
\tblue{The computations yield an overestimation of the average longitudinal velocity profile at the core of the channel which is representative of a slight excess in subgrid dissipation, leading to an underestimation of the friction at wall (and therefore of the friction velocity and the related Reynolds number). An underestimation of the spanwise and wall-normal velocity fluctuations is also observed, but much less pronounced for the CvP-Vreman computation, compared to the Vreman model alone. The CvP-Smagorinsky computation is in-between the CvP-Vreman and Vreman computations in terms of accuracy. The overall LES results are satisfactory given the coarse grid considered in this study. It is also seen that the CvP sensor is able to lower the intensity of the subgrid dissipation originating from the Vreman model yielding a better match with DNS for the mean and fluctuating velocity profiles. It is noteworthy that the CvP-Vreman approach yields accurate results both for free shear flows and wall-bounded turbulence without any tuning of the model parameter, making this model applicable to complex configurations featuring various physical phenomena.}

%% file: ChapelierS_JCP_2017.bbl
\begin{thebibliography}{10}
\newcommand{\enquote}[1]{``#1''}

\bibitem{smagorinsky1963general}
Smagorinsky, J., \enquote{General circulation experiments with the primitive
  equations,} {\em Mon. Weather Rev.\/}, Vol.~91, No.~3, 1963, pp.~99--164.

\bibitem{kraichnan1976eddy}
Kraichnan, R., \enquote{Eddy viscosity in two and three dimensions,} {\em J.
  Atmo. Sci.\/}, Vol.~33, No.~8, 1976, pp.~1521--1536.

\bibitem{sagautLESbook}
Sagaut, P., {\em Large-{E}ddy simulation for incompressible flows: an
  introduction\/}, Springer Verlag, 2006.

\bibitem{germano1991dynamic}
Germano, M., Piomelli, U., Moin, P., and Cabot, W., \enquote{A dynamic
  subgrid-scale eddy viscosity model,} {\em Phys. Fluids\/}, Vol.~3, No.~7,
  1991, pp.~1760--1765.

\bibitem{ghosal1995dynamic}
Ghosal, S., Lund, T.~S., Moin, P., and Akselvoll, K., \enquote{A dynamic
  localization model for {L}arge-{E}ddy simulation of turbulent flows,} {\em J.
  Fluid Mech.\/}, Vol.~286, 1995, pp.~229--255.

\bibitem{meneveau1996lagrangian}
Meneveau, C., Lund, T.~S., and Cabot, W.~H., \enquote{A Lagrangian dynamic
  subgrid-scale model of turbulence,} {\em J. Fluid Mech.\/}, Vol.~319, 1996,
  pp.~353--385.

\bibitem{salvetti1995priori}
Salvetti, M. and Banerjee, S., \enquote{A priori tests of a new dynamic
  subgrid-scale model for finite-difference large-eddy simulations,} {\em
  Physics of Fluids\/}, Vol.~7, No.~11, 1995, pp.~2831--2847.

\bibitem{lamballais1998spectral}
Lamballais, E., M{\'e}tais, O., and Lesieur, M., \enquote{Spectral-dynamic
  model for {L}arge-{E}ddy simulations of turbulent rotating channel flow,}
  {\em Theo. Comput. Fluid Dyn.\/}, Vol.~12, No.~3, 1998, pp.~149--177.

\bibitem{bou2005scale}
Bou-Zeid, E., Meneveau, C., and Parlange, M., \enquote{A scale-dependent
  Lagrangian dynamic model for large eddy simulation of complex turbulent
  flows,} {\em Phys. Fluids\/}, Vol.~17, No.~2, 2005, pp.~025105.

\bibitem{park2006dynamic}
Park, N., Lee, S., Lee, J., and Choi, H., \enquote{A dynamic subgrid-scale eddy
  viscosity model with a global model coefficient,} {\em Phys. Fluids\/},
  Vol.~18, No.~12, 2006, pp.~125109.

\bibitem{hughesLESVMS}
Hughes, T., Mazzei, L., and Jansen, K., \enquote{{Large-{E}ddy {S}imulation and
  the variational multiscale method},} {\em Computing and Visualization in
  Science\/}, Vol.~3, No.~1, 2000, pp.~47--59.

\bibitem{hugheschannelflow}
Hughes, T., Oberai, A., and Mazzei, L., \enquote{{Large-{E}ddy {S}imulation of
  turbulent channel flows by the variational multiscale method},} {\em Phys.
  Fluids\/}, Vol.~13, No.~6, 2001, pp.~1784--1799.

\bibitem{koobus2004variational}
Koobus, B. and Farhat, C., \enquote{{A variational multiscale method for the
  {L}arge-{E}ddy {S}imulation of compressible turbulent flows on unstructured
  meshes--application to vortex shedding},} {\em Comput. Meth. Appl. Mech.
  Eng.\/}, Vol.~193, No. 15-16, 2004, pp.~1367--1383.

\bibitem{farhat2006dynamic}
Farhat, C., Rajasekharan, A., and Koobus, B., \enquote{{A dynamic variational
  multiscale method for {L}arge-{E}ddy simulations on unstructured meshes},}
  {\em Comput. Meth. Appl. Mech. Eng.\/}, Vol.~195, No. 13-16, 2006,
  pp.~1667--1691.

\bibitem{bricteux2009multiscale}
Bricteux, L., Duponcheel, M., and Winckelmans, G., \enquote{A multiscale
  subgrid model for both free vortex flows and wall-bounded flows,} {\em Phys.
  Fluids\/}, Vol.~21, No.~10, 2009, pp.~105102.

\bibitem{wasberg2009variational}
Wasberg, C., Gjesdal, T., Reif, B., and Andreassen, {\O}., \enquote{Variational
  multiscale turbulence modelling in a high order spectral element method,}
  {\em J. Comput. Phys.\/}, Vol.~228, No.~19, 2009, pp.~7333--7356.

\bibitem{ouvrard2010classical}
Ouvrard, H., Koobus, B., Dervieux, A., and Salvetti, M., \enquote{Classical and
  variational multiscale {LES} of the flow around a circular cylinder on
  unstructured grids,} {\em Comput. Fluids\/}, Vol.~39, No.~7, 2010,
  pp.~1083--1094.

\bibitem{meyers2007evaluation}
Meyers, J. and Sagaut, P., \enquote{Evaluation of Smagorinsky variants in
  Large-Eddy simulations of wall-resolved plane channel flows,} {\em Phys.
  Fluids\/}, Vol.~19, No.~9, 2007, pp.~095105.

\bibitem{chapelier2016development}
Chapelier, J.-B., de~la Llave~Plata, M., and Lamballais, E.,
  \enquote{Development of a multiscale LES model in the context of a modal
  discontinuous Galerkin method,} {\em Comput. Meth. Appl. Mech. Eng.\/},
  Vol.~307, 2016, pp.~275--299.

\bibitem{stolz2005high}
Stolz, S., Schlatter, P., and Kleiser, L., \enquote{High-pass filtered
  eddy-viscosity models for {L}arge-{E}ddy simulations of transitional and
  turbulent flow,} {\em Phys. Fluids\/}, Vol.~17, 2005, pp.~065103.

\bibitem{vreman2003filtering}
Vreman, A., \enquote{The filtering analog of the variational multiscale method
  in Large-Eddy simulation,} {\em Phys. Fluids\/}, Vol.~15, No.~8, 2003,
  pp.~L61--L64.

\bibitem{jeanmart2007investigation}
Jeanmart, H. and Winckelmans, G., \enquote{Investigation of eddy-viscosity
  models modified using discrete filters: a simplified regularized variational
  multiscale model and an enhanced field model,} {\em Phys. Fluids\/}, Vol.~19,
  No.~5, 2007, pp.~055110.

\bibitem{david1993modelisation}
David, E., {\em Mod{\'e}lisation des {\'e}coulements compressibles et
  hypersoniques: une approche instationnaire\/}, Ph.D. thesis, 1993.

\bibitem{ackermann2001modified}
Ackermann, C. and M{\'e}tais, O., \enquote{A modified selective structure
  function subgrid-scale model,} {\em Journal of Turbulence\/}, Vol.~2, No.~1,
  2001.

\bibitem{chapelier2016spectral}
Chapelier, J.-B. and Lodato, G., \enquote{A spectral-element dynamic model for
  the Large-Eddy simulation of turbulent flows,} {\em J. Comput. Phys.\/},
  Vol.~321, 2016, pp.~279--302.

\bibitem{farge1999non}
Farge, M., Schneider, K., and Kevlahan, N., \enquote{Non-Gaussianity and
  coherent vortex simulation for two-dimensional turbulence using an adaptive
  orthogonal wavelet basis,} {\em Phys. Fluids\/}, Vol.~11, 1999, pp.~2187.

\bibitem{farge2001coherent}
Farge, M. and Schneider, K., \enquote{Coherent vortex simulation ({CVS}), a
  semi-deterministic turbulence model using wavelets,} {\em Flow Turb.
  Comb.\/}, Vol.~66, No.~4, 2001, pp.~393--426.

\bibitem{schneider2005coherent}
Schneider, K., Farge, M., Pellegrino, G., and Rogers, M.~M., \enquote{Coherent
  vortex simulation of three-dimensional turbulent mixing layers using
  orthogonal wavelets,} {\em J. Fluid Mech.\/}, Vol.~534, 2005, pp.~39--66.

\bibitem{farge2001coherent2}
Farge, M., Pellegrino, G., and Schneider, K., \enquote{Coherent vortex
  extraction in 3{D} turbulent flows using orthogonal wavelets,} {\em Phys.
  Rev. Letters\/}, Vol.~87, No.~5, 2001, pp.~054501.

\bibitem{rozema2015minimum}
Rozema, W., Bae, H.~J., Moin, P., and Verstappen, R.,
  \enquote{Minimum-dissipation models for large-eddy simulation,} {\em Phys.
  Fluids\/}, Vol.~27, No.~8, 2015, pp.~085107.

\bibitem{karamanos2000spectral}
Karamanos, G. and Karniadakis, G., \enquote{A spectral vanishing viscosity
  method for {L}arge-{E}ddy simulations,} {\em J. Comput. Phys.\/}, Vol.~163,
  No.~1, 2000, pp.~22--50.

\bibitem{pasquetti2006spectral}
Pasquetti, R., \enquote{Spectral vanishing viscosity method for {L}arge-{E}ddy
  simulation of turbulent flows,} {\em J. Sci. Comput.\/}, Vol.~27, No. 1-3,
  2006, pp.~365--375.

\bibitem{bogey2006large}
Bogey, C. and Bailly, C., \enquote{Large-{E}ddy simulations of transitional
  round jets: influence of the {R}eynolds number on flow development and energy
  dissipation,} {\em Phys. Fluids\/}, Vol.~18, No.~6, 2006, pp.~065101.

\bibitem{bogey2006computation}
Bogey, C. and Bailly, C., \enquote{Computation of a high {R}eynolds number jet
  and its radiated noise using {L}arge- {E}ddy simulation based on explicit
  filtering,} {\em Comput. Fluids\/}, Vol.~35, No.~10, 2006, pp.~1344--1358.

\bibitem{dairay2017numerical}
Dairay, T., Lamballais, E., Laizet, S., and Vassilicos, J.~C.,
  \enquote{Numerical dissipation vs. subgrid-scale modelling for Large Eddy
  simulation,} {\em J. Comput. Phys.\/}, Vol.~337, 2017, pp.~252--274.

\bibitem{grinstein2007implicit}
Grinstein, F.~F., Margolin, L.~G., and Rider, W.~J., {\em Implicit
  {L}arge-{E}ddy {S}imulation: computing turbulent fluid dynamics\/}, Cambridge
  university press, 2007.

\bibitem{thornber2007implicit}
Thornber, B., Mosedale, A., and Drikakis, D., \enquote{On the implicit large
  eddy simulations of homogeneous decaying turbulence,} {\em J. Comput.
  Phys.\/}, Vol.~226, No.~2, 2007, pp.~1902--1929.

\bibitem{hickel2006adaptive}
Hickel, S., Adams, N.~A., and Domaradzki, J.~A., \enquote{An adaptive local
  deconvolution method for implicit {LES},} {\em J. Comput. Phys.\/}, Vol.~213,
  No.~1, 2006, pp.~413--436.

\bibitem{hickel2007implicit}
Hickel, S. and Adams, N., \enquote{On implicit subgrid-scale modeling in
  wall-bounded flows,} {\em Phys. Fluids\/}, Vol.~19, No.~10, 2007, pp.~105106.

\bibitem{garnier1999use}
Garnier, E., Mossi, M., Sagaut, P., Comte, P., and Deville, M., \enquote{On the
  use of shock-capturing schemes for large-eddy simulation,} {\em J. Comput.
  Phys.\/}, Vol.~153, No.~2, 1999, pp.~273--311.

\bibitem{mittal1997suitability}
Mittal, R. and Moin, P., \enquote{Suitability of upwind-biased finite
  difference schemes for large-eddy simulation of turbulent flows,} {\em AIAA
  Journal\/}, Vol.~35, No.~8, 1997, pp.~1415--1417.

\bibitem{nagarajan2003robust}
Nagarajan, S., Lele, S., and Ferziger, J., \enquote{A robust high-order compact
  method for large eddy simulation,} {\em J. Comput. Phys.\/}, Vol.~191, 2003,
  pp.~392--419.

\bibitem{kravchenko1997effect}
Kravchenko, A. and Moin, P., \enquote{On the effect of numerical errors in
  {L}arge-{E}ddy simulations of turbulent flows,} {\em J. Comput. Phys.\/},
  Vol.~131, No.~2, 1997, pp.~310--322.

\bibitem{morinishi1998fully}
Morinishi, Y., Lund, T.~S., Vasilyev, O.~V., and Moin, P., \enquote{Fully
  conservative higher order finite difference schemes for incompressible flow,}
  {\em J. Comput. Phys.\/}, Vol.~143, No.~1, 1998, pp.~90--124.

\bibitem{leonard1974energy}
Leonard, A., \enquote{{Energy cascade in large-eddy simulations of turbulent
  fluid flows},} {\em Adv. Geophys.\/}, Vol.~18, 1974, pp.~237--248.

\bibitem{lesieur2001favre}
Lesieur, M. and Comte, P., \enquote{Favre filtering and macro-temperature in
  {L}arge-{E}ddy simulations of compressible turbulence,} {\em CR. Acad. Sci.
  Serie II\/}, Vol.~329, No.~5, 2001, pp.~363--368.

\bibitem{lesieur2005large}
Lesieur, M., M{\'e}tais, O., and Comte, P., {\em Large-{E}ddy simulations of
  turbulence\/}, Cambridge University Press, 2005.

\bibitem{lesieur1996new}
Lesieur, M. and Metais, O., \enquote{New trends in {L}arge-{E}ddy simulations
  of turbulence,} {\em Ann. Rev. Fluid Mech.\/}, Vol.~28, No.~1, 1996,
  pp.~45--82.

\bibitem{erlebacher1992toward}
Erlebacher, G., Hussaini, M., Speziale, C., and Zang, T., \enquote{Toward the
  {L}arge-{E}ddy simulation of compressible turbulent flows,} {\em J. Fluid
  Mech.\/}, Vol.~238, No.~1, 1992, pp.~155--185.

\bibitem{lele1992compact}
Lele, S., \enquote{Compact finite difference schemes with spectral-like
  resolution,} {\em J. Comput. Phys.\/}, Vol.~103, No.~1, 1992, pp.~16--42.

\bibitem{cook2005hyperviscosity}
Cook, A.~W. and Cabot, W.~H., \enquote{Hyperviscosity for shock-turbulence
  interactions,} {\em J. Comput. Phys.\/}, Vol.~203, No.~2, 2005, pp.~379--385.

\bibitem{persson2006sub}
Persson, P.-O. and Peraire, J., \enquote{Sub-cell shock capturing for
  discontinuous {G}alerkin methods,} {\em AIAA paper\/}, Vol.~112, 2006,
  pp.~2006.

\bibitem{metais1992spectral}
Metais, O. and Lesieur, M., \enquote{Spectral {L}arge-{E}ddy {S}imulation of
  isotropic and stably stratified turbulence,} {\em J. Fluid Mech.\/},
  Vol.~239, No.~1, 1992, pp.~157--194.

\bibitem{vreman2004eddy}
Vreman, A., \enquote{An eddy-viscosity subgrid-scale model for turbulent shear
  flow: {A}lgebraic theory and applications,} {\em Phys. Fluids\/}, Vol.~16,
  No.~10, 2004, pp.~3670--3681.

\bibitem{spyropoulos1996evaluation}
Spyropoulos, E.~T. and Blaisdell, G.~A., \enquote{Evaluation of the dynamic
  model for simulations of compressible decaying isotropic turbulence,} {\em
  AIAA J.\/}, Vol.~34, No.~5, 1996, pp.~990--998.

\bibitem{lilly1992proposed}
Lilly, D.~K., \enquote{{A proposed modification of the Germano subgrid-scale
  closure method},} {\em Physics of Fluids A: Fluid Dynamics (1989-1993)\/},
  Vol.~4, No.~3, 1992, pp.~633--635.

\bibitem{bagai1993flow}
Bagai, A. and Leishman, J.~G., \enquote{Flow visualization of compressible
  vortex structures using density gradient techniques,} {\em Exp. Fluids\/},
  Vol.~15, No.~6, 1993, pp.~431--442.

\bibitem{nemes2015mutual}
Nemes, A., Jacono, D.~L., Blackburn, H.~M., and Sheridan, J., \enquote{Mutual
  inductance of two helical vortices,} {\em J. Fluid Mech.\/}, Vol.~774, 2015,
  pp.~298--310.

\bibitem{quaranta2015long}
Quaranta, H.~U., Bolnot, H., and Leweke, T., \enquote{Long-wave instability of
  a helical vortex,} {\em J. Fluid Mech.\/}, Vol.~780, 2015, pp.~687--716.

\bibitem{coleman1995numerical}
Coleman, G., Kim, J., and Moser, R., \enquote{A numerical study of turbulent
  supersonic isothermal-wall channel flow,} {\em J. Fluid Mech.\/}, Vol.~305,
  1995, pp.~159--184.

\bibitem{kleckner2013creation}
Kleckner, D. and Irvine, W.~T., \enquote{Creation and dynamics of knotted
  vortices,} {\em Nature physics\/}, Vol.~9, No.~4, 2013, pp.~253--258.

\end{thebibliography}
